\documentclass[fleqn,10pt]{wlscirep}

\newcommand{\ud}[1]{{#1^{\dagger}}}
\newcommand{\bra}[1]{\left\langle #1\right|}
\newcommand{\ket}[1]{\left| #1\right\rangle}

\newcommand*{\triple}[2][.1ex]{%
  \mathrel{\vcenter{\offinterlineskip%
  \hbox{$#2$}\vskip#1\hbox{$#2$}\vskip#1\hbox{$#2$}}}}
\newcommand*{\triplerightarrow}{\triple{\rightarrow}}

\usepackage{amssymb}
\usepackage{amsmath,amsfonts,amssymb}
\usepackage{graphicx}
\usepackage{xcolor}
\usepackage{inputenc}

\title{Frequency-resolved Monte Carlo}

\author[1,2]{Juan Camilo L\'opez Carre\~no}
\author[1]{Elena del Valle}
\author[2,3,*]{Fabrice~P.~Laussy}

\affil[1]{Departamento de F\'isica Te\'orica de la Materia Condensada,
  Universidad Aut\'onoma de Madrid, 28049 Madrid, Spain}
\affil[2]{Faculty of Science and Engineering, University of Wolverhampton,
  Wulfruna St, WV1~1LY, United Kingdom}
\affil[3]{Russian Quantum Center, Novaya 100, 143025 Skolkovo,
  Moscow Region, Russia}

\affil[*]{f.laussy@wlv.ac.uk}

\begin{abstract}
  We adapt the Quantum Monte Carlo method to the cascaded formalism of
  quantum optics, allowing us to simulate the emission of photons of
  known energy. Statistical processing of the photon clicks thus
  collected agrees with the theory of frequency-resolved photon
  correlations, extending the range of applications based on
  correlations of photons of prescribed energy, in particular those of
  a photon-counting character.  We apply the technique to
  autocorrelations of photon streams from a two-level system under
  coherent and incoherent pumping, including the Mollow triplet regime
  where we demonstrate the direct manifestation of leapfrog processes
  in producing an increased rate of two-photon emission events.
\end{abstract}

\begin{document}

% \flushbottom
 \maketitle

\section{Introduction}
\label{sec:intro}  % \label{} allows reference to this section

The Monte Carlo method, of estimation by random sampling, was invented
by Ulam as a practical way to estimate the chance of winning a game of
solitaire, that a direct combinatorial approach had proved too
challenging for the then convalescing scientist who was playing cards
on a sick leave.~\cite{ulam_book91a} Since this had direct
applications for the more ``serious'' problems tackled at Los Alamos
(this was in 1946), such as the computation of neutron diffusion, the
activity had to receive a codename, which came after the Monaco ward,
famous for its casino.  The technique indeed relies on chance by
sampling randomly to get hold of a small but representative enough
sample to describe a system. This is a surprisingly powerful technique
that combines efficiency and accuracy, with applications in virtually
all fields of human endeavours.~\cite{knuth_book91a}

In quantum physics, the Quantum Monte Carlo technique finds many
ramifications in several fields.~\cite{ceperley86a} In the context of
interest in this work, that of quantum optics, several methods have
been developed in the early 90s (see Ref.~\cite{plenio98a} for a
review).  Of these, the quantum jump
approach~\cite{zoller87a,carmichael89b,dalibard92a,molmer96a} (see
Ref.~\cite{hegerfeldt09a} for an introduction) is particularly
appealing as it links the wavefunction collapse to the emission of a
photon. Assuming an ideal detector covering the full $4\pi$ solid
angle surrounding the emitter, this allows to perform a computer
experiment of photo-detections.  From such ``clicks'' (as we will call
a detected photon), one can for instance compute the Glauber
correlation functions~$g^{(n)}$ that measure the deviations of
intensity correlations at the~$n$th order from uncorrelated light, but
one can also compute less easily accessible quantities such as
exclusive probability densities, e.g., detecting the next photon at a
time~$\tau$ after one detection, with no other photon in between
($g^{(2)}$ assumes any photon rather than the next one), or
distributions of time delays between nearest neighbours, probabilities
to detect any given number or even configuration of photons in a time
window, or any other type of binning ``experiment''.  Such Quantum
Monte Carlo-generated photons have also been used to support the
introduction of the $N$-photon ``bundle'',~\cite{sanchezmunoz14a} to
distinguish the case of $N$-photon sources from strongly-correlated
emission at the~$N$ photon level.

In this text, we apply the quantum-jump Monte Carlo technique to the
case of filtered emission, that is to say, as applied to a stream of
photons going through an interference (i.e., Lorentzian) filter.  Mean
values for the correlators can be conveniently obtained with the
theory of frequency-resolved photon-correlations.~\cite{delvalle12a}
This theory predicts strong correlations in frequency windows that had
been neglected by both theorists and experimentalists until
recently,~\cite{gonzaleztudela13a, peiris15a} as they lie far away
from the luminescence peaks. Such correlations can clearly be turned
into a resource and when technology will be mature to exploit photons
as qubits, this aspect will certainly become compelling. As it should
be useful to go beyond mean values and get access to time series for a
variety of purposes, one could turn to the Quantum Monte Carlo
technique applied to color-resolved photons. While the quantum Monte
Carlo method has been used to compute the power spectrum as well as
time-series of photon
emission,~\cite{dum92a,marte93a,molmer93a,garraway95a,hegerfeldt96a,plenio96a,schack96a}
its combined use for both time \emph{and} energy-resolved photons has,
to the best of our knowledge, not been provided before in both a
practical and exact form, as the few attempts in this
direction~\cite{tian92b,sanchezmunoz14b} have involved a cavity in the
weak-coupling limit, which comes at the price of vanishing signal or
approximate correlations if coupling of the filter is not weak
enough. Monte Carlo methods have been used instead for their
computational advantage or to access particular configurations such as
the resonance fluorescence spectrum of an effective-three level system
in a bright period exclusively (such systems are noted for their
intermittent emission).  In contrast, our approach allows to extract
streams of photons from any frequency windows of a quantum source,
using all the signal theoretically available and taking into account
self-consistently the effect of its filtering, with an exact treatment
of its effect on the correlations. This allows to revisit photon
counting experiments with the added energy degree of freedom, that are
already challenging without the frequency constrains.

In this text, we first prove that the technique is exact and,
subsequently, we cover two cases for illustration. Namely, we show the
effect of filtering a two-level system, and describe how this spoils
the antibunching and quantum character of such sources in a practical
context, although from a theoretical point of view, the saturated
emitter is the brightest single-photon emitter, but of photons with
wildly fluctuating frequencies (so not indinstinguishable). We
illustrate how the loss of antibunching in time deviates notably from
the single-exponential approximation used in the
literature.~\cite{arXiv_delvalle17a}.  We compare both the cases of
coherent and incoherent excitations at low pumping. Then, at high
pumping in the coherent case, thus bringing the problem in the Mollow
triplet regime, we show how this turns a simple system into a
versatile, tunable quantum source, with applications such as quantum
spectroscopy~\cite{lopezcarreno15a,mukamel15a,kazimierczuk15a} or
photon sources with tunable statistics.~\cite{dory17a} We will also
address arguments~\cite{shatokhin16a} that imply that the strongly
correlated emission is an artifact of normalization, which we will
rebute by explicitly exhibiting these strongly correlated photons
thanks to the Monte Carlo simulations.  Before that, however, we
briefly summarize the theory of frequency-resolved photon correlations
and its main conclusions, which are to be found in greater details
elsewhere,~\cite{delvalle12a,gonzaleztudela13a,lopezcarreno15a,delvalle13a,sanchezmunoz14b,gonzaleztudela15a,sanchezmunoz15a,lopezcarreno16a,lopezcarreno16b}
and prove that the proposed Monte Carlo method scheme is
Mathematically equivalent to this exact theory.

\section{Theory}

Although the process of filtering the light emitted from an optical
source has a clear interpretation---the emitted photons are detected
in a certain frequency window---its theoretical description used to be
far from trivial.~\cite{vogel_book06a} With the introduction of the
sensor method,~\cite{delvalle12a} this became a straightforward task
no more complicated than any old problem of computing quantum
correlators, getting rid of all the complicated tasks of
normal-ordering and time-integrals in spaces of many dimensions.  The
technique relies on coupling the system to sensors with strength
$\varepsilon$ and taking the limit of vanishing coupling. In such a
limit, it is enough to consider only two levels of the sensors as
their populations remain $\ll1$, without affecting the system’s
dynamics.  The computation of normalized correlators results in
quantities that are~$\varepsilon$ independent to first order and exact
in the limit $|\varepsilon|\rightarrow0$ (taking, in general,
$\varepsilon \in \mathbb{C}$).  These results are also absolute in the
sense that they do not depend on detection efficiency or other details
of the measurement, but characterize the source's emission in given
frequency windows.  Alternatively,~\cite{flayac14a} such a coupling
can also be made through the so-called ``cascaded
formalism'',~\cite{gardiner93a,carmichael93a} that describes the
dynamics of ``detectors'', which are physical objects with a sizable
coupling to the source (unlike sensors that have vanishing coupling)
but that also do not alter the dynamics of the source, regardless of
how strongly they are affected by it. Each method presents some
advantages: the sensor method is straightforward to implement while
the cascading formalism allows to characterize the detector's dynamics
beyond normalized correlations. An important difference with respect
to computational cost is that while the sensor is typically described
by a two-level system, the detector must be described by an harmonic
oscillator, since it does get populated, whereas the sensor only acts
as a probe in the limit of vanishing coupling. As a result, the sensor
always provides exact results as a two-level system while the detector
must be truncated high enough to provide a close-enough approximation,
which depends on the dynamics of the system, and is therefore a tricky
question. Furthermore, when considering cross-correlations, instead of
$N$ two-level systems, one is dealing with $N$ harmonic oscillators
and the problem becomes numerically forbidding, while the sensors'
Hilbert space scales as $2^N$ which is still tractable for $N$ much
larger than anything that has been considered so far
experimentally. For autocorrelations of the $N$th order with the
sensor method, one can also use an harmonic oscillator truncated
to~$N$ excitations instead of $N$ two-level systems degenerate in
frequency, which is also exact and with no need of checking
convergenve for higher truncations, as is the case of the cascaded
formalism. So in cases where correlations are requested rather than an
actual signal, we believe the sensor method to be preferable, as it is
both more efficient and more robust. In the present text, however, we
specifically require a signal and will therefore turn to the cascaded
formalism.

The mathematical equivalence of the two approaches for normalized
autocorrelations can be established as follows. On the one hand, the
sensor method ``plugs'' sensors to the dynamics. Formally,
calling~$\sigma$ the annihilation operator of a source and $\xi$ that
of a sensor probing it, we can describe their joint dynamics by a
Liouvillian equation
\begin{equation}
  \label{eq:Thu6Jul171246BST2017}
  \partial_t \rho = i [\rho,H_\sigma+\omega_\xi\ud{\xi}\xi+\varepsilon \ud{\sigma}\xi+\varepsilon^* \sigma\ud{\xi}] +
  \frac{\gamma_\sigma}{2} \mathcal{L}_\sigma \rho +
  \frac{\gamma_\xi}{2}  
  \mathcal{L}_\xi \rho\,,
\end{equation}
where $H_\sigma$ is the Hamiltonian that describes the internal
dynamics of the source, $\gamma_\sigma$ is the decay rate of the
source, $\gamma_{\xi}$ is the decay rate of the sensor and
$\mathcal{L}_c\rho \equiv 2c\rho \ud{c} - \ud{c}c\rho - \rho \ud{c}c$.
The dynamics of an arbitrary operator
$\xi^{\dagger\mu}\xi^{\nu}\sigma^{\dagger m}\sigma^n$ under the action
of this Liouvillian is described with the notations of
Ref.~\cite{delvalle12a}, by the equation (we set~$\hbar=1$ along the
paper)
\begin{equation}
  \label{eq:Thu6Jul163739BST2017}
  \partial_t\vec w[\mu,\nu]=\large\{ M+[(\mu-\nu)i\omega_\xi-(\mu+\nu){\frac{\gamma_\xi}2}]\mathbf{1} \large\}  \vec w[\mu,\nu]+i\varepsilon \mu
  T_+\vec w[\mu-1,\nu] -i \varepsilon^* \nu T_- \vec w[\mu,\nu-1]+O(\varepsilon^2)\,,
\end{equation}
where $\mathbf{1}$ is the unit matrix, $T_\pm$ are normal-ordering
superoperators for the $\sigma$ operators and $\vec w[\mu,\nu]$ is a
vector of correlators for the $\mu$th and~$\nu$th powers of the sensor
operators~$\xi$, $\xi^\dagger$ and spanning in normal order all powers
of the~$\sigma$, $\sigma^\dagger$ operators, i.e.,
$\vec w[\mu,\nu]\equiv(\langle\xi^{\dagger\mu}\xi^\nu\rangle,
\langle\xi^{\dagger\mu}\xi^\nu\sigma\rangle,
\langle\xi^{\dagger\mu}\xi^\nu\ud{\sigma}\rangle,\cdots,\langle\xi^{\dagger\mu}\xi^\nu
\sigma^{\dagger m}\sigma^n\rangle,\cdots)^T$.
The $O(\epsilon^2)$ notation means that all other terms are of higher
order than~$\varepsilon$.  The matrix $M$ provides the dynamics for
the source, $\partial_t\vec w[0,0]=M \vec w[0,0]+O(\epsilon^2)$, and
is independent of the sensor at the lowest order in~$\varepsilon$. At
this stage, we do not assume any property of~$\sigma$ or $\xi$, which
could be bosonic (in which case $\mu$, $\nu$, $m$ and~$n$ are
unbounded) or fermionic (in which case $\mu$, $\nu$, $m$ and~$n$ are 0
or~1). Equation~(\ref{eq:Thu6Jul163739BST2017}) can be integrated,
which yields
\begin{equation}
  \label{eq:MonNov27143432CET2017}
  \vec w[\mu,\nu]=i |\varepsilon| \left\{ M+[(\mu-\nu)i\omega_\xi-(\mu+\nu){\frac{\gamma_\xi}2}]\mathbf{1} \right\}^{-1} \left(-e^{i\theta}\mu
  T_+\vec w[\mu-1,\nu] + e^{-i\theta} \nu T_- \vec w[\mu,\nu-1] \right)+O(\varepsilon^2)\,,
\end{equation}
where $\varepsilon=|\varepsilon|e^{i\theta}$. This in turn can be
solved recursively, down to $w[0,0]$ where the equation
self-truncates. This provides $\mu\nu+\mu+\nu$ terms, all with the
same leading order of $\varepsilon^{\mu+\nu}$.  More details on this
derivation can be found in the Supplementary Material of
Ref.~\cite{delvalle12a}.  The important point is that normalised
correlators are ratios of components of $\vec w[\mu,\nu]$ with the
$(\mu+\nu)$th power of components of $\vec w[1,1]$, themselves of
order~$\varepsilon$, so that such ratio cancel out $\varepsilon$ to
leading order. While the higher-order terms do not have to cancel
exactly, as is the case of the leading-order terms, they become
negligible as the sensor coupling is made smaller, which does not
preclude, however, a finite result for the normalized correlators,
obtained from the $\varepsilon$ independent term, in contrast to
unnormalised correlators that vanish with $\varepsilon$. Therefore, in
the limit $\varepsilon\rightarrow0$, the result becomes exact.

On the other hand, the cascaded formalism, which aims at exciting a
target without affecting the source, provides a similar type of
cancellation, although not restricted to vanishing coupling.  From a
causality point of view, it is clear that such a source/detector
scenario where only one affects the other can be realized. The source
that emitted a photon towards a detector may not even exist anymore by
the time the detector is excited.  This is achieved formally through
interferences that cancel the back-action from the detector to the
source. The master equation describing this asymmetric coupling reads
\begin{equation}
  \label{eq:SunApr23163037BST2017}
   \partial_t \rho = i [\rho,H_\sigma+\omega_\xi \ud{\xi}\xi] +
  \frac{\gamma_\sigma}{2} \mathcal{L}_\sigma \rho +
  \frac{\gamma_\xi}{2}  
  \mathcal{L}_\xi \rho + \sqrt{\alpha \gamma_\sigma \gamma_\xi} \lbrace
  [\sigma \rho, \ud{\xi}] + [\xi,\rho\ud{\sigma}] \rbrace\,.
\end{equation}
The last three-terms of
Eq.~(\ref{eq:SunApr23163037BST2017}) can be re-written in the Lindblad
form as
\begin{equation}
  \label{eq:SunApr23165942BST2017}
  \frac{\gamma_\sigma}{2} \mathcal{L}_\sigma \rho +
  \frac{\gamma_\xi}{2}  
  \mathcal{L}_\xi \rho + \sqrt{\alpha \gamma_\sigma \gamma_\xi} \lbrace
  [\sigma \rho, \ud{\xi}] + [\xi,\rho\ud{\sigma}] \rbrace = 
  \frac{1}{2}\mathcal{L}_{\hat{o}} \rho +
  \frac{\chi_1\gamma_\sigma}{2} 
  \mathcal{L}_\sigma\rho + \frac{\chi_2\gamma_\xi}{2}\mathcal{L}_\xi \rho + \frac{\sqrt{\alpha \gamma_\sigma
  \gamma_\xi}}{2} [\rho ,\ud{\xi}\sigma-\ud{\sigma}\xi]\,,
\end{equation}
where $\hat{o}=\sqrt{(1-\chi_1) \gamma_\sigma}\,\sigma +
\sqrt{(1-\chi_2) \gamma_\xi}\,\xi$ is the joined decay operator of the
whole system, source and detector, and the interpretation of the
factor~$\chi_k$ becomes that of factors that quantify the amount of
signal that each part, source and detector, generates on its own and
that the joined system generates as a whole. The detector, which must
have a finite lifetime to couple to the source, thus also has an
intrinsic frequency window with effect of filtering the emission it
detects, whence the connection to the sensors formalism. The
factor~$\alpha=(1-\chi_1)(1-\chi_2)$, for~$0\le\chi_1,\chi_2\le1$,
takes into account that the source can have several decay
channels. This is required for instance when only fluorescence is
wanted without contamination from another source, e.g., a laser
exciting it (experimentally this is typically achieved by detecting at
right angle from the exciting beam).  

Our proof proceeds by showing that
$\xi^{\dagger\mu}\xi^{\nu}\sigma^{\dagger m}\sigma^n$ has the same
equation as in the sensor formalism, by computing explicitly the
equation for~$\partial_t\vec w[\mu,\nu]$ in the cascaded formalism,
Eq.~(\ref{eq:SunApr23165942BST2017}). This reads, to all orders in the
coupling in this case:
\begin{equation}
  \label{eq:Thu6Jul173801BST2017}
  \partial_t\vec w[\mu,\nu] =\large\{ M+[(\mu-\nu)i\omega_\xi-(\mu+\nu){\frac{\gamma_\xi}2}]\mathbf{1} \large\}  \vec w[\mu,\nu]-\sqrt{\alpha \gamma_\sigma \gamma_\xi} \{\mu T_+\vec w[\mu-1,\nu]+\nu
  T_-\vec w[\mu,\nu-1]\}\,.
\end{equation}
Remarkably, this equation has the same form as
Eq.~(\ref{eq:Thu6Jul163739BST2017}) with $\varepsilon \rightarrow i
\sqrt{\alpha \gamma_\sigma \gamma_\xi}$.  Even though $\varepsilon$ is
complex and a vanishing quantity in
Eq.~(\ref{eq:Thu6Jul163739BST2017}), with higher order corrections,
and $\sqrt{\alpha \gamma_\sigma \gamma_\xi}$ it real and finite in
Eq.~(\ref{eq:Thu6Jul173801BST2017}), both methods provide exactly the
same normalised correlators, as these coupling parameters enter in
both the numerators and denominators with the same power and cancel
out.  The result becomes exact for vanishing coupling in the case of
sensors and is exact in all cases with the cascaded formalism,
regardless of their normalisation. Note as well that $\theta$, the
phase of the coupling $\varepsilon$, has an effect on the dynamics
only if the Lindblad equation features products of different operators
in its dissipative terms, which is the case for the cascaded formalism
with~$\mathcal{L}_{\hat{o}}$ that brings cross terms of~$\sigma$
and~$\xi$. The sensor formalism, however, has no such joint decay
emission and the phase of~$\varepsilon$ does not play any role, so
that $\varepsilon$ could have beeen set real.  This achieves to prove
the mathematical equivalence of the sensor method with the cascaded
formalism for the computation of normalized correlators.

Since the sensor formalism has been shown\cite{delvalle12a} to be
equivalent to normalized photon correlations according to
photo-detection theory,~\cite{vogel_book06a} the above equivalence of
the sensor and cascaded formalisms shows that applying the quantum
Monte Carlo method to the detector\cite{carmichael_book08a} realises a
sampling of the emission in the corresponding frequency windows, from
which one can reconstruct the frequency-resolved photon
correlations. That is to say, this allows us to simulate the photon
emission with both time and energy information, which is what we are
going to illustrate in the following. Note that with both
Eqs.~(\ref{eq:Thu6Jul163739BST2017})
and~(\ref{eq:Thu6Jul173801BST2017}), any given
correlator~$\langle\xi^{\dagger\mu}\xi^\nu\rangle$ can be computed
exactly (by recurrence) in terms of lower order ones
$\langle\xi^{\dagger \mu'}\xi^{\nu'}\rangle$ only, with
$\mu' \leq \mu$ and $\nu'\leq \nu$. This means that both methods can
be applied using $N$ two-level systems as detectors at different
frequencies, in order to compute cross correlations, or with a single
harmonic oscillator truncated at $N$ excitations, to compute the $N$th
order monocromatic autocorrelation function $g^{(N)}$. This is however
not sufficient with the cascaded formalism for computing the density
matrix (full state) of the detectors or for doing Monte Carlo
simulations of the emission. In such cases, one must model the $N$
detectors as harmonic oscillators with a high enough truncation to
provide converged results. The simulation is conveniently implemented
through the quantum-jump approach. The dynamics of the system is thus
described by a wavefunction~$\ket{\psi(t)}$ that occasionally
undergoes a process of ``collapsing'', attributed to the emission of a
photon, that one records in the simulation as a detector would
register a click in an experiment. The collapse is decided in each
infinitesimal time interval~$\delta t\rightarrow0$, where the
evolution of the wavefunction is governed by two elements: a
non-Hermitian Hamiltonian and random quantum jumps. In a system
described by the master equation
$\partial_t \rho = i[\rho,H] + (1/2)\sum_k \mathcal{L}_{c_k}\rho$, the
non-Hermitian Hamiltonian is constructed as
$\tilde H = H - (i/2)\sum_k \ud{c_k}c_k$, and the evolution of the
wavefunction is given either by
\begin{equation*}
%  \label{eq:SunApr23183331BST2017}
\ket{\psi(t+\delta t)} =  \frac{c_k \ket{\psi(t)}}{\bra{\psi(t)}
                             \ud{c_k}c_k \ket{\psi(t)}}
                           \quad \mathrm{or} \quad
   \ket{\psi(t+\delta t)} =  \frac{\exp(i \tilde H \delta t)
    \ket{\psi(t)}}{ \bra{\psi(t)} \exp(i \ud{\tilde H} \delta t)
                             \exp(i \tilde H \delta t) \ket{\psi
                               (t)}}\,,
\end{equation*}
depending on whether the system undertook or not a quantum jump,
respectively.  The probability that this happened in the
interval~$\delta t$ due to the operator~$c_k$ is proportional to the
mean value of this operator, namely
$p_k= \bra{\psi}\ud{c_k}c_k\ket{\psi} \delta t$.  For the system
described by the master equation~(\ref{eq:SunApr23163037BST2017}), the
collapse operators are
\begin{equation}
  \label{eq:SunApr23184131BST2017}
 c_1 = \sqrt{(1-\chi_1) \gamma_\sigma}\,\sigma +
   \sqrt{(1-\chi_2) \gamma_\xi}\,\xi \,, \quad c_2 = \sqrt{\chi_1
    \gamma_\sigma} \,\sigma\, \quad \mathrm{and} \quad c_3 =
  \sqrt{\chi_2 \gamma_\xi}\, \xi\,,
\end{equation}
whereas the non-Hermitian Hamiltonian is given by
\begin{subequations}
  \begin{align}
    \tilde H &= H_\sigma + H_\xi -\frac{i}{2}\sqrt{\alpha \gamma_\sigma
               \gamma_\xi} (\ud{\xi}\sigma -\ud{\sigma}\xi) - \frac{i}{2}
               (\ud{c_1}c_1 + \ud{c_2}c_2 + \ud{c_3}c_3)\,,\\
             &= H_\sigma + H_\xi -  i \sqrt{\alpha \gamma_\sigma
               \gamma_\xi}\, \ud{\xi}\sigma- \frac{i}{2}( \gamma_\sigma \ud{\sigma}\sigma + \gamma_\xi \ud{\xi}\xi) \,.
  \end{align}
\end{subequations}
The Monte Carlo approach within the cascaded formalism context has
already been considered~\cite{carmichael93a} but without connecting it
to frequency filtering. Now that we have established this
correspondence mathematically, we illustrate in the following how the
clicks collected through
Eqs.~(\ref{eq:SunApr23163037BST2017}-\ref{eq:SunApr23165942BST2017}),
in the frequency window determined by the detector, match indeed with
the correlations predicted by the theory of frequency-resolved photon
correlations.~\cite{delvalle12a}  Here, we apply this technique to the
driven two-level system, under both coherent and incoherent emission,
at low and large pumping. While some of the underlying physics has
been published elsewhere, this will allow us to revisit it from
another angle and provide additional results.  We will consider both
the cases of autocorrelations and cross-correlations. Although the
same principles could be extended to even more than two detectors, we
postpone such discussions and their further applications to future
works.

\subsection{The two-level system}

The simplest system in quantum physics is the two-level system. We
find it inspiring that it is still a Research problem, constantly
offering new questions. Manifestly, in a quantum universe, the
two-level system is as complicated as anything else. We first discuss
a very basic problem, namely, the effect of filtering on photon
emission from the Monte Carlo point of view. Starting with incoherent
pumping, at rate~$P_\sigma$, with Hamiltonian
\begin{equation}
  \label{eq:Fri28Apr194221BST2017}
  H_\sigma=\omega_\sigma\ud{\sigma}\sigma
\end{equation}
for a two-level system with free energy~$\omega_\sigma$, and
Liouvillian
$\partial_t \rho = i[\rho,H_\sigma]+ (\gamma_\sigma/2)
\mathcal{L}_\sigma \rho + (P_\sigma/2) \mathcal{L}_{\ud{\sigma}}\rho$,
with~$\gamma_\sigma$ the inverse lifetime, one finds a simple enough
dynamics of Glauber's second order correlator:
\begin{equation}
  \label{eq:Fri28Apr194345BST2017}
  g^{(2)}(\tau)=1-\exp[-(\gamma_\sigma+P_\sigma)\tau]\,,
\end{equation}
\begin{figure}[t]
  \centering
  \includegraphics[width=\linewidth]{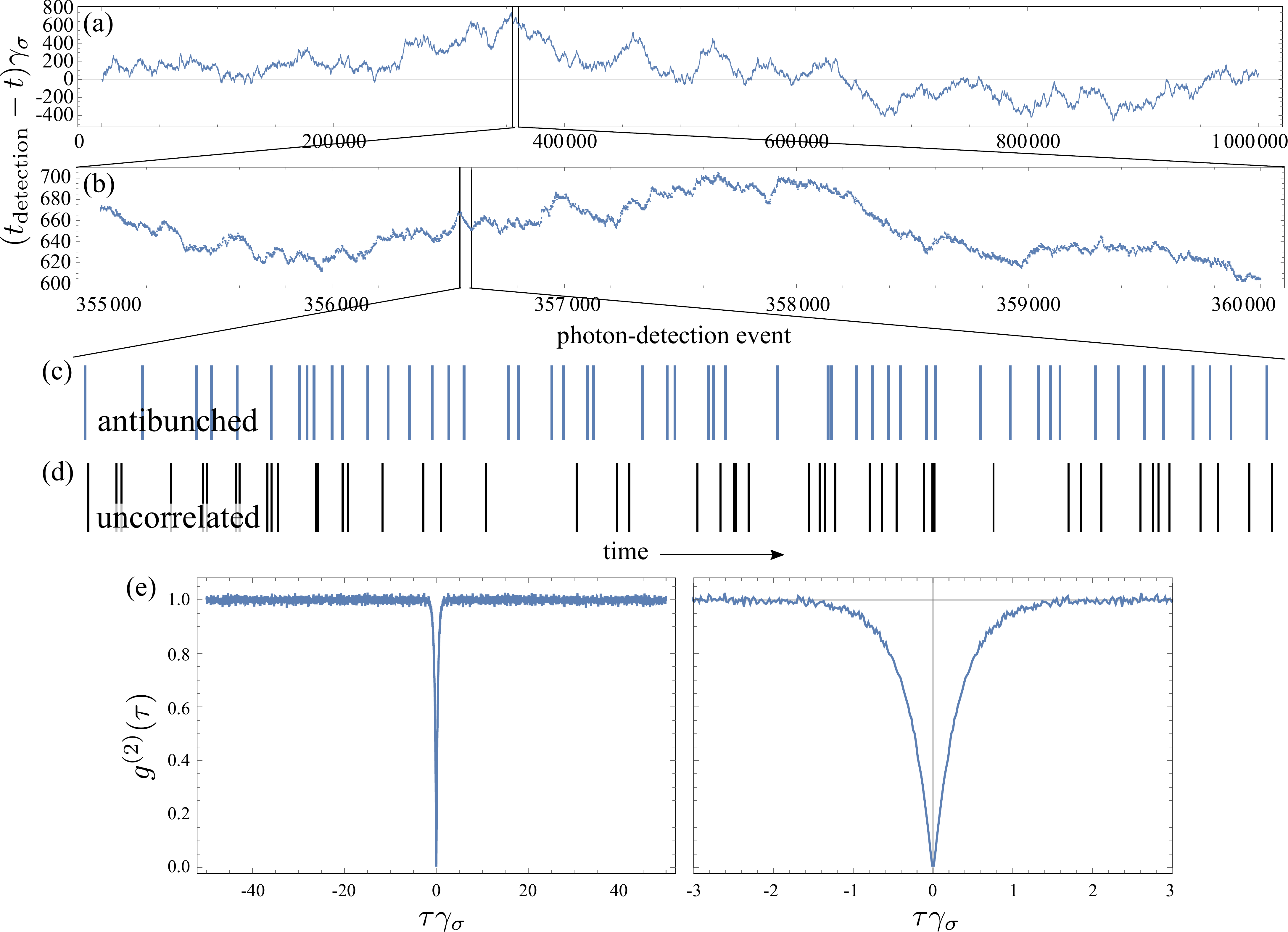}
  \caption{Monte Carlo method on a two-level system. (a) Times of
    emission for 1\,000\,000 recorded photons as compared to their
    mean emission rate, exhibiting a classical random walk. (b) Zoom
    of~(a) in the highlighted window. (c) Zoom of~(b) in the
    highlighted window, with detected photons now displayed in
    absolute time rather than relatively to their mean emission
    time. Locally, one can observe a structure in their statistical
    distribution, with a tendency of ordering and mutual
    repulsion. This becomes obvious when comparing with uncorrelated
    photons with the same emission rate, shown in (d). The latter
    exhibit Poisson bursts. (e) Intensity correlations~$g^{(2)}(\tau)$
    computed from the one million points, in two timescales, featuring
    a clear antibunching. The inverse lifetime of the two-level
    system sets the time unit.}
  \label{fig:Sat29Apr193603BST2017}
\end{figure}
with, in particular, $g^{(2)}(0)=0$, that is, perfect antibunching. A
(conventional) Monte Carlo simulation using the technique explained
above, is shown in Fig.~\ref{fig:Sat29Apr193603BST2017}. The upper
panel shows the fluctuations in the detection times of a million
photons from such a source. As such, this realizes a random walk,
similar to a random (Poissonian) process, and at large timescales
there is nothing noticeable. On the short timescale, however, one can
observe clear correlations of antibunching, as shown in the series of
clicks indicated by blue ticks in
Fig.~\ref{fig:Sat29Apr193603BST2017}(c). Namely, photons tend to repel
each other and appear more orderly than if they would be uncorrelated,
as is the case of the second series of photon detections, shown for
comparison with black ticks in
Fig.~\ref{fig:Sat29Apr193603BST2017}(d). The uncorrelated series
exhibits the counter-intuitive ``Poisson clumping'' or ``Poisson
burst'' effect,~\cite{feller_book68a} made famous by von Bortkiewicz's
horse kicking casualties in the Prussian army and still of recurrent
appearance in the medias as intuition repels the notion that a burst
of accidents in, say, a hospital, is a natural random process rather
than negligence. The strongly-correlated character of the two-level
system emission becomes clear and compelling when computing intensity
correlations $g^{(2)}(\tau)$ from the clicks, defined as the density
of probability of finding two photons with a time
difference~$\tau$. Specifically, from the times of detection~$t_i$, we
compute $t_i-t_j$ for all $1\le i\le N$ with $N$ the total number of
detected photons (here~$N=10^6$) and compare the density of time
differences to that from uncorrelated clicks with the same
intensity. Note that in a typical experiment, a first photon starts a
timer and a second stops it, and a distribution of the time difference
between successive photons is used as a good approximation. In our
case, we compute the exact correlations by collecting all the time
differences within the correlation window of interest. This is shown
for $|\tau|\le50/\gamma_\sigma$ in
Fig.~\ref{fig:Sat29Apr193603BST2017}(e), left. One sees an overall
plateau, indicating that photons have the same distribution for
long-time separations as if they were emitted by a Poisson process
(randomly). But one also observes a clear dip at~$\tau\approx0$,
indicating that at such close distances, photons behave very
differently than uncorrelated ones, namely, the occurrence of small
time delays is strongly suppressed.  This is better resolved in
Fig.~\ref{fig:Sat29Apr193603BST2017}(e), right. Such a behaviour
defines antibunching, $g^{(2)}(0)<g^{(2)}(\tau)$, with coincidences,
i.e., simultaneous detection of two photons, less likely to occur than
other closely spaced detections, with perfect suppression of
coincidences when $g^{(2)}(0)=0$. Since these correlations wash out at
long times, one has $\lim_{\tau\rightarrow\infty}g^{(2)}(\tau)=1$. The
time it takes to reach this plateau is the second-order coherence
time.  We do not overlap these results of the Monte Carlo signal with
the theory curve, Eq.~(\ref{eq:Fri28Apr194345BST2017}), since, with
one million points, it is exact to within the plot accuracy. Beside
the statistical noise, that starts to be apparent
for~$\tau>1/(2\gamma_\sigma)$, the Monte Carlo data provides a smooth
curve in the window of strong correlations. In our simulation, the
$\Delta t$ was $0.01/\gamma_\sigma$ and the binning size was taken
twice as large, corresponding to the two closely-spaced vertical lines
on the right panel of Fig.~\ref{fig:Sat29Apr193603BST2017}(e),
bounding~$g^{(2)}(0)$ from below due to this small uncertainty.  With
a binning size equal to the Monte Carlo timestep, one recovers the
perfect antibunching at the origin, although on two grid points, so
also producing a small error (the result would be perfect only in the
limit of vanishing timesteps).

These results provide the background for our approach in the filtered
signal. The general question is, what happens to the emitted photons
if a filter is interposed on their way to the detector? This does not
simply subtract a fraction, it also redistributes those that make it
through, to provide them with possibly very different statistical
properties, as we now discuss in more details.

\section{Emission of a filtered two-level system}

\begin{figure}[th]
  \centering
  \includegraphics[width=\linewidth]{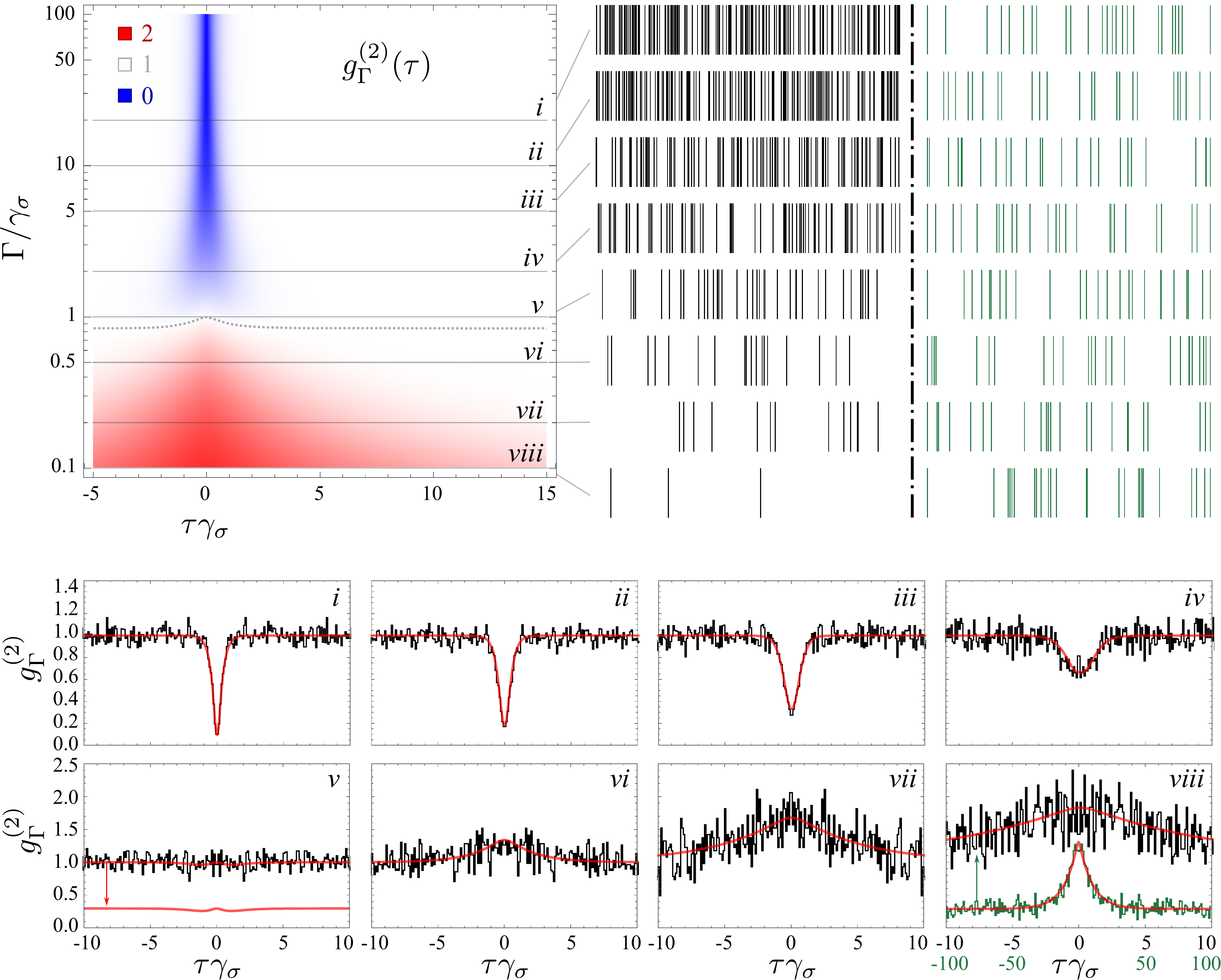}
  \caption{Frequency-resolved emission from an incoherent driven
    two-level system.  The number of events (clicks) recorded are
    close to 10\,000 (namely 9976, 9916, 9974, 9927, 9967, 9955, 9860)
    for the cases~$i$--$vii$ respectively, and 25\,000 for case~$viii$
    to get enough signal for the small timescale comparison to the
    other filters. The density-plot is the theoretical
    $g^{(2)}_\Gamma(\tau)$ with the color code indicated (blue for
    antibunching, red for bunching and white for
    uncorrelated). Filtering leads to thermalization. The transition
    is slightly more complex than merely loss of antibunching. The
    dotted line shows the isoline $g^{(2)}_\Gamma(\tau)=1$. Monte
    Carlo simulations have been done for the eight cuts shown. Samples
    of clicks are shown in the same time window (black, left) or with
    rescaling to have the same intensity (green, right). There is a
    neat transition visible to the naked eye between the two types of
    photon statistics. Autocorrelation computed from the clicks are
    shown in the eight panels at the bottom, together with the theory
    prediction. In panel~v, the theory curve is also shown displayed
    to reveal its fine structure departing from
    $g^{(2)}_\Gamma(\tau)=1$. In panel~viii, also the case of longer
    times is shown since thermalization goes together with slowing
    down of the dynamics. For the density plot $1/\gamma_\sigma$ sets
    the unit and $P_\sigma=2\gamma_\sigma$.}
  \label{fig:Sat29Apr131334BST2017}
\end{figure}

\subsection{Incoherent excitation}

The effect of a Lorentzian filter on the statistics of emission of an
incoherently excited two-level system is shown in
Fig.~\ref{fig:Sat29Apr131334BST2017}.  The theory predicts
thermalization and loss of antibunching with narrowing filtering. The
exact way how this happens is discussed
elsewhere,~\cite{arXiv_delvalle17a} and the theoretical result is
shown on the density plot in Fig.~\ref{fig:Sat29Apr131334BST2017}
along with eight Monte-Carlo simulations of roughly 10\,000 clicks
each (25\,000 for the narrowest filter in case~$viii$). Extracts of
the recorded clicks are shown, comparing them 1) in the same time
window (black ticks), with effect of having much less clicks for
narrower filters, and also 2) when rescaling the unit of time so that
the intensities are the same (green ticks). In the latter case, one
can compare the statistical distributions, and observe the transition
from antibunched clicks~($i$) to thermal ones~($viii$) passing by
auxiliary distributions. In the former case, one observes the
characteristic antibunching, equally-spaced like distribution of a
two-level emitter. In the latter case, one finds the wildly
fluctuating thermal (or chaotic) light, with pronounced bunching in
the form of long gaps of no emission followed by gusts of
emission. This can be differentiated even with the naked eye from the
Poisson distribution, whose tendency for ``clumping'' does not get as
dramatic as the thermal case. One can follow the transition neatly
from these various sets of clicks, passing by the case of uncorrelated
light.

Since the isoline~$g^{(2)}_\Gamma(\tau)=1$ is not straight~
\cite{arXiv_delvalle17a} (it is shown as a dotted line in the density
plot of Fig.~\ref{fig:Sat29Apr131334BST2017}), the passage from
antibunching to bunching does not transit through exactly uncorrelated
(or coherent light), although the deviation is too small to be
appreciated on a small sample. To observe such fine variations, one
needs to acquire a large statistical ensemble and condense the
correlations in a single object, such as~$g^{(2)}_\Gamma$, as is shown
in the eight panels at the bottom of
Fig.~\ref{fig:Sat29Apr131334BST2017}. The case~$v$ of
close-to-uncorrelated light is also shown separately from the Monte
Carlo data to reveal its fine structure. The other cases have a
simpler shape of a dip that turns into a hump. The correlation time
also changes dramatically, as is observable both from the density plot
and the Monte Carlo histograms. As the emission thermalizes, its
fluctuations occur on longer timescales. This is the reason for the
increased noise in panels~$vi$--$viii$. There, one should increase the
binning and consider larger time windows, as shown in green for
case~$viii$ that assumes a binning of $\Delta t\gamma_\sigma=1$
instead of $0.1$ for the other cases, and plot the correlations in a
time window $|t\gamma_\sigma|\le100$ instead of 10, as indicated on
the respective axes, recovering the excellent agreement with theory
displayed by the antibunched cases.

Now in possession of the statistical data, and with the insurance of
its accuracy given its agreement with the theory, it is possible to
undertake various types of analysis that would not be so
straightforward theoretically, as has been described in the
introduction.  We will not go in this direction now and leave for
future works and/or other colleagues (the statistical data is
available on a public repository). Instead, we now turn to the case of
coherent excitation, that also presents features of interest.

\subsection{Coherent excitation}

\begin{figure}[t]
  \centering
  \includegraphics[width=\linewidth]{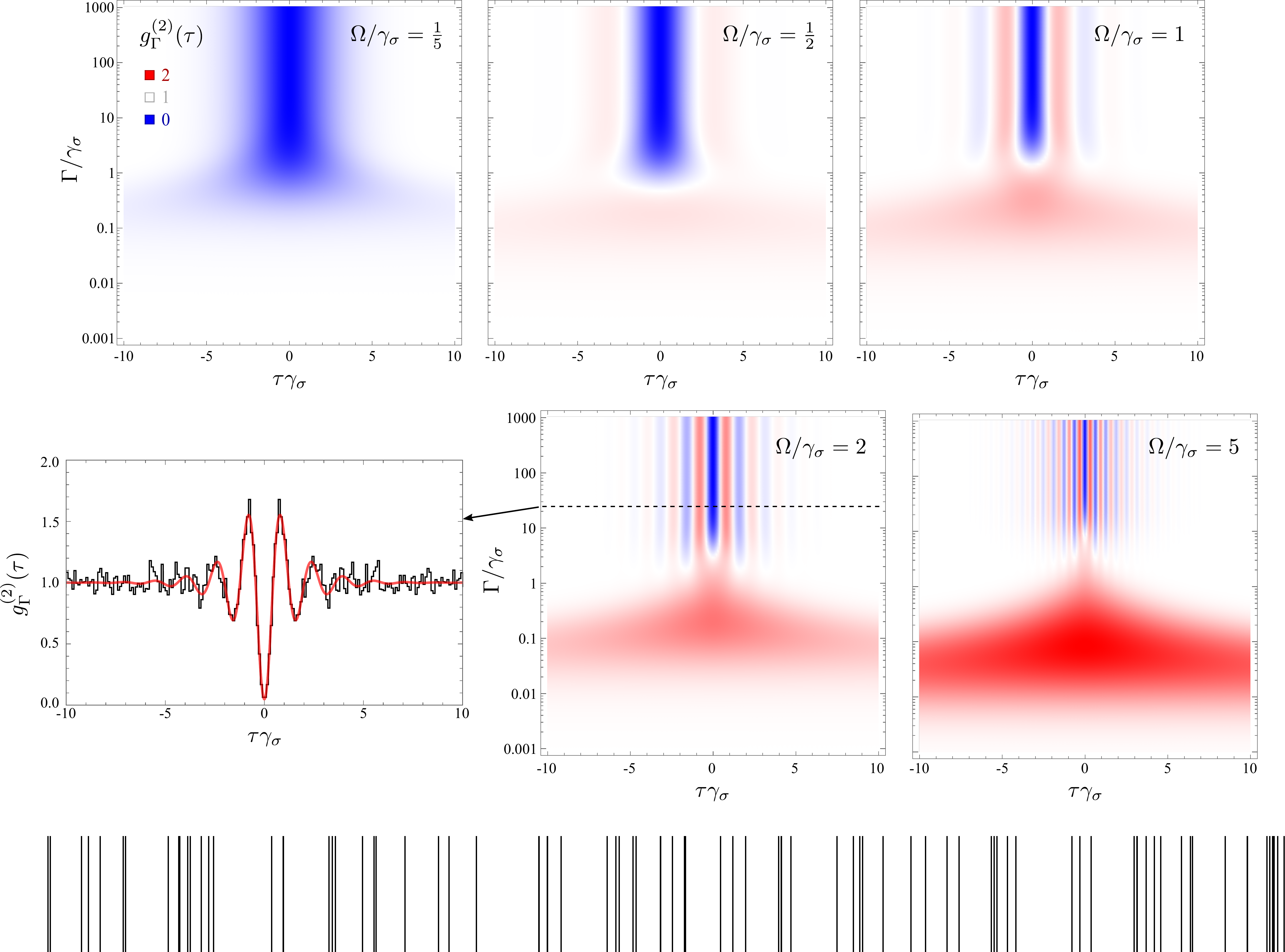}
  \caption{Frequency and time-resolved $g^{(2)}_\Gamma(\tau)$ of a
    two-level system coherently driven, in its transition from the
    Heitler to the Mollow regime (from left to right). At low pumping,
    one does not observe thermalization (bunching) with narrowing
    filters. Higher pumping brings both bunching, similar to the case
    of incoherent pumping, and oscillations. The bunching is observed
    only for moderately narrow filtering as extremely narrow filtering
    goes back to filtering exclusively the Rayleigh peak, with a
    resurgence of the Heitler effect and uncorrelated (or coherent)
    emission. Wide filtering overlapping the three peaks captures the
    Rabi oscillations. A Monte Carlo simulation of the case
    highlighted with the dashed line is shown through a small sample
    of clicks (bottom) and the autocorrelation function, compared to
    the theory prediction. There is a clear structure in the photon
    clicks, that is unlike any of the cases shown previously.}
  \label{fig:Sat29Apr135840BST2017}
\end{figure}

The case of filtering coherent is shown in
Fig.~\ref{fig:Sat29Apr135840BST2017}. Here as well, there is
thermalization, although this occurs in the case of
strong-driving,~\cite{arXiv_delvalle17a} it is interesting to
consider the effect of filtering and approach it from the Monte Carlo
perspective. Taking one slice featuring these oscillations, we collect
$10^5$ clicks, a small portion of which is shown as ticks at the
bottom of Fig.~\ref{fig:Sat29Apr135840BST2017}.  Computing the
autocorrelations, we find indeed strong oscillations from a very good
antibunching with steep bunching elbows, in agreement with the
theory. This produces even more pronounced correlations in the
photon-detection events, where the spacing appears more regular and
between clumps of photons. As far as continuous streams are concerned,
this suggests that such strongly-oscillating $g^{(2)}$ do in fact
provide more ordered time series than the conventional antibunching of
the type of Eq.~(\ref{eq:Fri28Apr194345BST2017}). Such questions are
however beyond the scope of the present text. We conclude this Section
with further comments on the Heitler effect (coherence of the Rayleigh
peak), that is broken at high pumping, but is eventually restored with
narrow-enough filtering. First, regarding the emergence of a
thermalization similar to that of incoherent driving,
cf.~Fig.~\ref{fig:Sat29Apr131334BST2017}, this is obtained when one
enters the Mollow triplet regime.~\cite{mollow69a}  In this case,
luminescence has split into a triplet lineshape and, when filtering at
resonance (as is the case here), one filters the central peak alone,
which is known to correspond to the spontaneous emission of a photon
that leaves the state of the dressed two-level system
unchanged.~\cite{reynaud83a} As such, the spontaneously emitted
photons react to filtering in a similar way than the incoherently
pumped two-level system, hence the observed bunching for narrowing
filters linewidths. The similarity is only partial, however, as
instead of thermalization, with $g^{(2)}_\Gamma(0)=2$, the transition
is to a super-chaotic state, with $g^{(2)}_\Gamma(0)=3$ in the limit
of infinite pumping~\cite{gonzaleztudela13a} (for the parameters
considered here, we find $\max_\Gamma g^{(2)}_\Gamma(0)\approx2.2$).
More strikingly, when filtering well within the central peak, one then
isolates the Rayleigh ($\delta$) peak again and reverts to the
low-pumping case, with the statistics becoming uncorrelated, as shown
in Fig.~\ref{fig:Sat29Apr135840BST2017}. Large filtering windows, on
the other hand, collect the emission from all three peaks and
reproduce the Rabi oscillations, which is the case selected for the
Monte Carlo sampling. We explore in more details the opportunities
offered by the Mollow triplet in
Section~\ref{sec:Sun30Apr135221BST2017}.

\subsection{Effective quantum state}
\label{sec:Sun30Apr113440BST2017}

\begin{figure}[t]
  \centering
  \includegraphics[width=.75\linewidth]{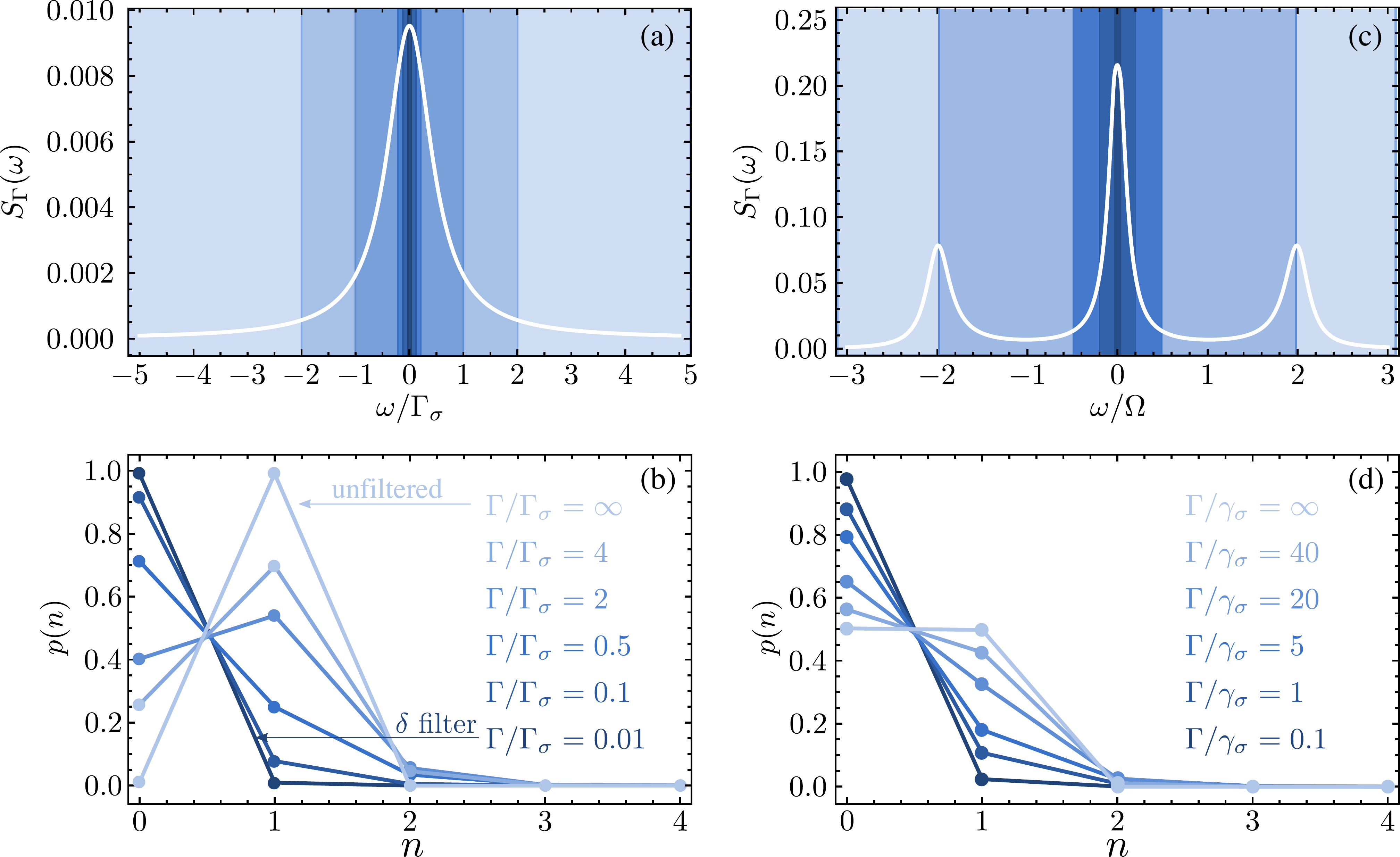}
  \caption{Effective quantum state reconstruction from the emission of
    a two-level system under high (a) incoherent and (b) coherent
    excitation. The various blue areas correspond to different filters
    linewidths. The light filtered in this way is equivalent to an
    unfiltered emitter whose quantum state has distribution~$p(n)$ to
    have~$n$ excitations, and is shown in (b) and~(d),
    respectively. For the incoherently driven two-level system,
    case~(a), the system is kept in its excited state by large
    pumping, so that without filtering, one observes a state close to
    the Fock state~$p(n)=\delta_{n,1}$. Filtering leads to
    thermalization, with preponderance of vacuum but nonzero
    probability to detect~$n> 1$ particles. The same is observed for
    the coherently driven two-level system, case~(b), but starting
    from~$p(0)=p(1)=1/2$ due to the no-inversion of a two-level system
    in presence of stimulated emission.  Parameters: for incoherent
    excitation, $P_\sigma=10^2\gamma_\sigma$ (saturating the two-level
    system). For coherent excitation, $\Omega=5\gamma_\sigma$. The
    rest of the parameters are as indicated in the figure.}
  \label{fig:Sat29Apr193837BST2017}
\end{figure}

In this Section, we adapt the method of Zubizarreta Casalengua
\emph{et al.}~\cite{zubizarretacasalengua17a} to get the photon
distribution~$p(k)$ of a frequency-resolved emitter, i.e., the
probabilities for it to have~$k$ photons, which are, equivalently, the
diagonal elements of its effective density matrix. Namely, the
frequency-filtered source is regarded as an effective source of its
own. Since filtering typically turns the original source into one of
another kind, the effective source gets attributed a new annihilation
operator~$s$ rather than the original~$\sigma$. As we consider a
photon source and we know that regardless of the original emitter, its
filtering can produce an arbitrary number of photons, we assume
that~$s$ is bosonic. The detector and the effective source are of
course related. On the one hand, the normalized correlators are
identical, since the detector measures faithfully the correlations of
the source:
\begin{equation}
  \label{eq:Tue30May161556BST2017}
  g^{(n)}=\frac{\langle s^{\dagger n}s^n\rangle}{\langle\ud{s}s\rangle^n}=\frac{\langle \xi^{\dagger n}\xi^n\rangle}{\langle\ud{\xi}\xi\rangle^n}\,,
\end{equation}
and as, on the other hand, there is conservation of energy (assuming
an ideal detector):
\begin{equation}
  \label{eq:Tue30May161810BST2017}
  \gamma_\sigma\langle\ud{s}s\rangle=\Gamma\langle\ud{\xi}\xi\rangle\,,
\end{equation}
since the rate of emission from the effective source is also that of
detection. We find from combining
Eqs.~(\ref{eq:Tue30May161556BST2017}) and~(\ref{eq:Tue30May161810BST2017})
the relation between the unnormalized correlators:
\begin{equation}
  \label{eq:Sun30Apr223743BST2017}
  \langle s^{\dagger n}s^n\rangle=\left(\frac{\Gamma}{\gamma_\sigma}\right)^n\langle\xi^{\dagger n}\xi^n\rangle\,,
\end{equation}
The statistics~$p(n)$ of the photon emission from the effective source
can now be obtained by inverting the relation
$\langle s^{\dagger n}s^n\rangle=\sum_{k=n}^N({k!}/{(k-n)!})p(k)$
(with~$N$ large enough) to provide the probabilities~$p(k)$ for the
effective source to have~$k$ quanta of excitation, for integer~$k$. In
the cascaded formalism, the correlators of the detectors,
$\langle\xi^{\dagger n}\xi^n\rangle$ can be computed from the master
equation, and are source dependent. The expressions for the population
of a detector being fed by an incoherently and coherently driven
two-level system are given by L\'opez Carre{\~no} and
Laussy~\cite{lopezcarreno16a} at resonance. Here, we provide a more
general version for the detector at an arbitrary frequency.  The
incoherent case reads\footnote{Note that there is a typo in
  Ref.~\cite{lopezcarreno16a} with an extra factor~4; the correct
  result is as given here.}
\begin{equation}
  \label{eq:Sun30Apr212744BST2017}
  \langle\xi^{\dagger}\xi\rangle = \frac{P_\sigma
      (P_\sigma+\Gamma+\gamma_\sigma)}{(P_\sigma +
      \gamma_\sigma) ((P_\sigma+\Gamma+\gamma_\sigma)^2+\Delta^2)}
\end{equation}
and the coherent excitation reads
\begin{multline}
  \label{eq:Sun30Apr213040BST2017}
  \langle\xi^{\dagger}\xi\rangle = \Big[2\gamma_{01}\Omega^2
    (\gamma_{10}(\gamma_{11}^2+4\Delta^2)^2 (\gamma_{12}^2+4\Delta^2)+
    4 \Omega^2 (\gamma_{10}\gamma_{11}^2 \gamma_{12}\gamma_{32}+
    4(2\gamma_{10}^3 + 16\gamma_{10}^2
    \gamma_{01}+23\gamma_{10}\gamma_{01}^2+8\gamma_{01}^3)\Delta^2 \\
    -
    16( \gamma_{10}-2\gamma_{01})\Delta^4) + 32
    \gamma_{11}(\gamma_{10}^2+ 4\Delta^2)\Omega^4) \Big]\bigg/ \Big[
    (\gamma_{10}^2+4\Delta^2)
  (\gamma_{11}^2+4\Delta^2)(\gamma_{01}^2+4\Omega^2)\times\nonumber\\
  \lbrace\gamma_{10}^4
    +6 \gamma_{10}^3 \gamma_{01} + 12
    \gamma_{10}\gamma_{01}(\gamma_{01}^2+2\Delta^2+4\Omega^2)
    +
    \gamma_{10}^2[13\gamma_{01}^2+8(\Delta^2+2\Omega^2)] +
    4[ \gamma_{01}^4+
    4(\Delta^2-2\Omega^2)^2+\gamma_{01}^2(5\Delta^2+8\Omega^2)]\rbrace\Big]
\end{multline}
with $\gamma_{ij}\equiv i\Gamma+j\gamma_\sigma$.  From a Monte Carlo
simulation in a time~$T$, the detector population is obtained as the
ratio between the total number of clicks recorded
and~$\gamma_\sigma T$. This allows to obtain the luminescence spectrum
by scanning the detector in frequency (we do not show it but have
checked it to be the case). One can now reconstruct the diagonal
elements of the effective density matrix that, under an unspecified
dynamics, is seen through the detector to yield the recorded
photo-detection events.  Since this can be achieved from (all) the
Glauber correlators and the knowledge of the emitter's mean population
(known from the radiative lifetime), one can recover the effective
$p(n)$ in this way.  This allows us to access new classes of quantum
steady states, tailored by frequency-filtering. We now illustrate how
this takes shape in the case already discussed of filtered two-level
system emission, starting with the case of incoherent excitation.
This is shown in Figs.~\ref{fig:Sat29Apr193837BST2017}(a--b) for the
filtered saturated two-level system, i.e., where the system is held in
its excited state by very large pumping, $P_\sigma\gg\gamma_\sigma$,
so that the density matrix reads $p(0)\rightarrow0$
and~$p(1)\rightarrow1$. With Eq.~(\ref{eq:Sun30Apr212744BST2017}) or
from the detected clicks, one can compute the population and
reconstruct this quantum state of the emitter, namely, the Fock
state~$p(n)=\delta_{n,1}$ ($\delta$ being the Kronecker function). The
application of a filter turns the system from a two-level emitter to a
source able to deliver more than one photon at a time, namely, for
narrow enough filtering, $p(n)\approx(1-\theta)\theta^n$ for all~$n$,
with~$\theta\approx0.01$ for the narrowest filter considered here. We
have lost two orders of magnitude for the population but the
probability to observe two (resp.~three) particles is 1\%
(resp.~0.01\%) of that to observe one only, etc., which effectively
shows how filtering a single-photon source turns it into a black body
with nonzero probability to emit $n$ photons.  Its population is
smaller than without the filter, as the latter is rejecting some
photons, but the statistical distribution of those that go through now
corresponds to an altogether different quantum state. A similar
situation occurs with coherent excitation (when not filtering so much
as to isolate the Rayleigh peak), with, for the case shown in
Fig.~\ref{fig:Sat29Apr193837BST2017}(c,d), $\theta\approx0.025$. In
both cases, one can see in this way at which point filtering prevents
a single-photon source to emit non-classical states of
light,~\cite{arXiv_lopezcarreno16c} for instance by comparing $p(1)$
to $3\sqrt{3}/(4e)\approx47.8\%$, the smallest probability above which
a state is non-Gaussian.~\cite{filip11a}

\section{Tuneable statistics from the Mollow triplet}
\label{sec:Sun30Apr135221BST2017}

\subsection{Auto-correlations}

\begin{figure}[th]
  \centering
  \includegraphics[width=\linewidth]{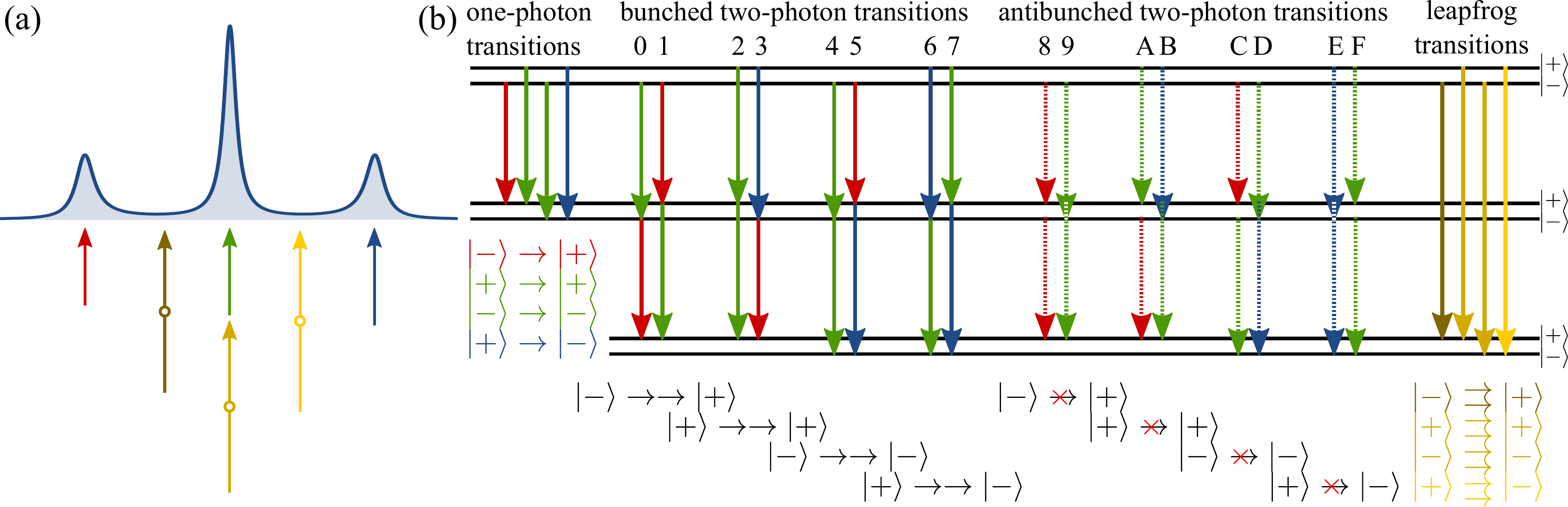}
  \caption{(a) Line shape and (b) level structure of the Mollow
    triplet. The lineshape is easily understood as the result of
    one-photon transitions between neighbour rungs of the
    dressed-state ladder, shown in red, green and blue. Since there
    are two degenerate green arrows, the central peak to which they
    correspond is twice as large. This picture is also helpful to
    visualise two-photon transitions. All the possible combinations
    are shown in~(b). Combinations such as~8 and~E, that occur only
    once, behave as expected. Other cases that occur multiple times,
    such as red and green transitions that happen in 0, 1, A and C,
    require exact computations to identify their actual
    behaviour. Also shown are the ``leapfrog transitions'', that jump
    over the intermediate manifold in a direct two-photon
    emission. The cases where photons have the same frequency is shown
    in panel~(a).}
  \label{fig:Fri26May193749BST2017}
\end{figure}

The Mollow regime that splits the luminescence into three lines
provides in this way new natural spectral windows. One of the obvious
questions this brings forward is: what is the statistics of the
photons emitted by the three peaks? (is it the same as the total
emission? One could imagine that all three peaks are emitting
antibunched light since they ultimately originate from a two-level
system. Our discussions so far in simpler systems should prepare us to
find otherwise). The triplet structure, first computed exactly by
Mollow~\cite{mollow69a} but without providing a physical picture as to
its origin, can actually be well understood from a simple model,
introduced by Cohen-Tannoudji \emph{et al.}: the ``dressed
atom''.~\cite{cohentannoudji77a} In this model, the combined
atom+laser is considered as a new entity, with a new structure of
energy levels, shown in Fig.~\ref{fig:Fri26May193749BST2017}, and in
which the transitions between the states account for the
photoluminescence. On the basis of this picture, by considering
two-photon transitions, one can foresee some correlations between the
peaks.  The
transitions~$\ket{p}\rightarrow\ket{\pm}\rightarrow\ket{q}$ for
any~$p,q=+$ or~$-$, that go down the Mollow ladder, could be expected
to result in bunching.  In contrast, since one cannot chain in this
way $\ket{p}\rightarrow\ket{-}$ and~$\ket{+}\rightarrow\ket{q}$ or
$\ket{p}\rightarrow\ket{+}$ and~$\ket{-}\rightarrow\ket{q}$, one can
expect antibunching.  Inspection of all the combinations of two-photon
relaxations ``suggests'' that:
\begin{itemize}
\item each side peak is antibunched (cases~8 and~E in
  Fig.~\ref{fig:Fri26May193749BST2017}),
\item the central peak comes with both bunching~(2 and~4) and
  antibunching (9 and~F),
\item the central peak comes with both bunching (0, 1, 6, 7) and antibunching
  (A, B, C, D) with each side peaks,
\item the side peaks are bunched together (3, 5).
\end{itemize}
(One could also go further and consider time-ordering and/or detuning.)
Early calculations by Apanasevich and Kilin~\cite{apanasevich79a} and
Cohen Tannoudji and Reynaud~\cite{cohentannoudji77a} confirmed the
side-peaks antibunching in autocorrelation and their mutual bunching
in cross-correlations. In a later work with more involved
calculations, Schrama \emph{et al.}~\cite{schrama91a} have shown that
the central and side peaks feature no mutual correlations. This is due
to interferences in the order of emission, that could nevertheless be
linked to the co-existence of bunched and antibunched emission
events. This shows that while intuition is strongly supported by the
dressed-atom picture, it does not dispense from exact calculations for
cases that could be ambiguous (and it confirms that there is indeed
agreement with expectations for cases that cause no ambiguity, such as
side-peaks emission).

With the theory of frequency-resolved photon
correlations,~\cite{delvalle12a} it is straightforward to compute such
correlations exactly, like Mollow did for the luminescence spectrum,
without referring to the dressed-states structure.  This also allows to
consider cases ouside their frequency windows, in fact, the complete
landscape of two-photon correlations can be
obtained.~\cite{gonzaleztudela13a,delvalle13a}  In the case of the
Mollow triplet,~\cite{gonzaleztudela13a,lopezcarreno17a} it
shows that the triplet structure reverberates at the two-photon level,
through the apparition of a set of 3 hyperplanes, that obey the
``leapfrog'' equations
\begin{equation}
  \label{eq:FriMar17153039GMT2017}
  \omega_1 +\omega_2=\Delta\,,\quad \mathrm{with}
  \quad  \Delta=-\Omega_+,0,\Omega_+\,.
\end{equation}
(The same applies at the~$N$-photon
level~\cite{lopezcarreno17a}). The triplet structure comes, at
any photon-number level, from the three possibilities to join the two
dressed initial and final states. Note that while there are four
transitions, $N$ is assumed sufficiently large for two transitions to
be degenerate.  The name of ``leapfrog'' comes from the fact that, at
the $N$-photon level, transitions can occur by jumping over $N-1$
intermediate manifolds.  This relaxes energy conservation for the
individual photons and restricts it to the combined emission. In this
sense, this is a ($N$ photons) version of G\"opper Mayer's two-photon
processes.~\cite{goeppertmayer31a} that is however typically difficult
to access (a notable exception is the planetary nebulae
continuum~\cite{spitzer51a}). The case~$N=2$ is shown on the right of
Fig.~\ref{fig:Fri26May193749BST2017}. In the case of resonance
fluorescence, such correlations have been noted by Apanasevich and
Kilin~\cite{apanasevich77a} for
the~$\ket{\pm}\rightrightarrows\ket{\pm}$ case (that is, overlooking
the $\ket{\pm}\rightrightarrows\ket{\mp}$ counterparts, which are
equally obvious with the dressed atom picture in mind). Interestingly,
part of this school of researchers, who has produced noteworthy works
on the problem of photon
correlations,~\cite{shatokhin00a,shatokhin01a,shatokhin02a} has
recently expressed some critics on this leapfrog picture, writing
that~\cite{shatokhin16a} ``\emph{the concept of the ``leapfrog''
  processes is not justified}'', that they ``\emph{present an
  alternative explanation}'' based on ``\emph{the unnormalized
  spectral correlation function}'' which is, they write, ``\emph{a
  true measure of spectral correlations}'' and ``\emph{which exhibits
  no signatures of the leapfrog transitions}''. From their discussion,
one thus understands that the production of strongly-correlated
photons away from the peaks, that we predict, is an artifact due to
normalization.  In principle, one can indeed inflate the vacuum and
create what one could regard as an artificial superbunching. This is
not what happens with leapfrog emission, however, although according
to these authors, nothing of interest takes place away from the peaks,
what they illustrate by producing a two-photon spectrum remarkably
featureless, in contrast to our two-photon spectrum that is rich from
photon correlations flourishing away from the peaks.  Their
non-normalized spectrum is correct but, we believe, is also not
interesting as it merely shows that first order processes smother
second-order ones, as is however well-known and expected. We show, in
contrast, that the scarce signal from higher-order processes has
stronger correlations than those from first-order processes. This will
be amply and vividly illustrated through Monte Carlo simulations
below. One can also consider placing a cavity in this ``featureless''
region when not normalized, and observe how the system then keeps
emitting strongly correlated photons but now dominating over the other
first-order processes,~\cite{sanchezmunoz14a,arXiv_sanchezmunoz17a}
which would not happen would the correlation be an artifact due to
normalization. We will show below thanks to the frequency-resolved
Monte Carlo simulations how one can anyway see the manifestation of
leapfrog processes with the naked eye.
\begin{figure}[t]
  \centering
  \includegraphics[width=.94\linewidth]{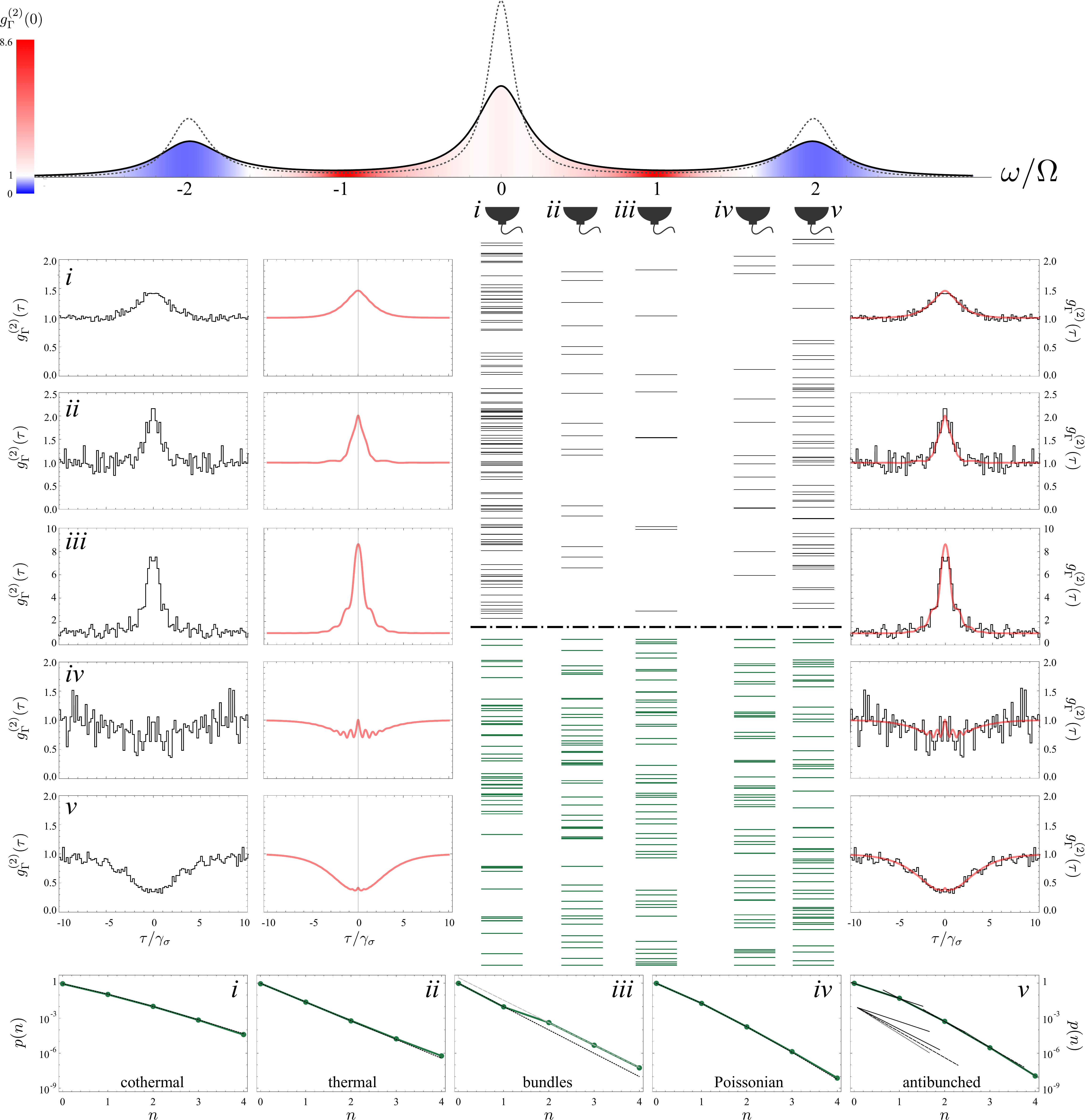}
  \caption{Frequency-resolved Monte Carlo simulation of the Mollow
    triplet in independent frequency windows. The dotted (solid)
    lineshape is the triplet as detected by an ideal
    (finite-bandwidth~$\Gamma$) detector. Sequences of clicks in the
    frequency windows~$i$--$v$ have been recorded, with 17241, 22836,
    99457, 9112 and 46126 events, respectively. Small samples are
    displayed, in the same time window (black ticks, up) or with
    renormalization of time to compare equal intensities (green ticks,
    bottom). Clear structures are visible even to the naked eye, in
    particular, the existence of leapfrog emission is obvious. Only
    one detector has been used, so the streams are not here
    cross-correlated. The autocorrelations are shown as measured by
    the Monte Carlo data (left column), as computed by the theory of
    frequency-resolved photon correlation (2nd column, red) and both
    superimposed (right column), to show their rich fine structure and
    the overall agreement.  The effective quantum state reconstruction
    is shown at the bottom, together with fits to fundamental
    distributions. Panel~$v$ has its successive emission probabilities
    brought together to show the exponential extinction of higher
    photon-numbers. $1/\gamma_\sigma$ sets the unit and
    $\Omega=5\gamma_\sigma$.}
  \label{fig:Thu27Apr184800BST2017}
\end{figure}
The rest of their discussion is only semantics, in which case we
should clarify, as this is apparently needed, that the dressed-atom
picture is, precisely, a ``picture'', that is, an insightful mental
representation that is helpful to visualize the basic mechanism at
play, support the intuition and guide further inquiries. This does not
preclude exact calculations based on the opaque equations of quantum
mechanics. This is possibly why the Mollow triplet is named after
Benjamin, not only for his seminal input but also in recognition of
the exact expression, although the Cohen Tannoudji and Reynaud
approximate picture is the one everybody has in mind when thinking
about this problem. We combine both approaches, the sensor technique
provides the exact result, while the leapfrog processes provide a
physical representation \emph{\`a la} Cohen Tannoudji \emph{et
  al.}. Thus, in the same way that resonance fluorescence is not
spontaneous emission from the dressed atom, the leapfrog emission is
not, strictly speaking, spontaneous emission jumping over intermediate
manifolds. This is, instead, a complicated process that involves the
laser and the two-level system in a sequence of coherent absorption
and emission. We have in fact shown~\cite{lopezcarreno16b} how even
fluorescence in the low-driving regime does not consist of Rayleigh
scattering events but form an intricate interference between emission
and absorption, that powers the single-photon emission mechanism by
suppressing the fat tails of the Lorentzian and turning the lineshape
into a~$t_2$ distribution instead. In the strongly nonlinear regime, a
similar dynamics takes place but it becomes forbidding and certainly
not even useful to apprehend the problem in these terms. Note that
although inspired by the dressed-atom approach of the problem,
ultimately, our computations are exact (and in full agreement with
this physical picture).  While the dressed atom has proven to be
extremely fruitful for their purpose, we find it to be even more so in
our $N$-dimensional case,~\cite{lopezcarreno17a} if not
mandatory. It is, indeed, very easy based on this concept to
understand why some configurations have less strong correlations than
others, for instance, one $\ket{+}\triplerightarrow\ket{+}$ transition
with one photon non-degenerate with the two others (of interest for
photon-heralding of two-photon emission), is particularly weak. This
is because it is resonant with the
$\ket{-}\rightarrow\ket{+}\rightrightarrows\ket{-}$ transition, that
breaks this channel of relaxation by interposing a real state in the
three-photon emission. Configurations
$\ket{+}\triplerightarrow\ket{+}$ whose energies do not intersect with
real states, on the other hand, have very strong correlations and are
more suitable for heralding purposes. It would seem difficult to make
sense of these observable facts, that follow from exact computations,
without the leapfrogging concept. In fact we can easily generalise
them to arbitrary photon orders and guess which energies are to be
avoided in a~$N$ dimensional space to harness the best sought
configuration of multi-photon emission, without having to undertake
any actual computation. Is the concept therefore not justified and
should one only be allowed infinite series of Feynman diagrams? We
believe that the comments of Shatokhin \emph{et al.} targetting the
leapfrog picture bring very little to the discussion, if not in fact
muddying it with confused statements and blurring their actual
technical contributions that otherwise concur with our results, and of
which we wish to remove nothing, as this approach has its merits.

Back to the general discussion. We show in
Fig.~\ref{fig:Thu27Apr184800BST2017} the statistics of clicks from
photo-detection events of the Mollow triplet in frequency windows
spanning from the central peak to the side peaks, including various
other windows in between, in particular, the leapfrog window. Note
that, here as well, the data is a single-detector observable, that is
to say, the different streams shown are not correlated to each others
as they have been obtained by the same detector in different runs of
the experiment. It would require~$5$ detectors to obtain the same
result but with correct cross-correlations (this is beyond the scope
of the present discussion that will go up to two detectors only, but
is of course a topic of interest for applications). As we did in
Fig.~\ref{fig:Sat29Apr131334BST2017}, we show both ticks in a given
time window (in black) and with a rescaling of the unit of time so
that their densities are equal (in green). Here as well, the relative
emission rates mean that longer integration times are required when
collecting away from the peaks. The gain in terms of correlation
strengths, however, makes it worthwhile to focus on these regions of
reduced emission, in a spirit akin to distillation: trading quantity
for quality. The frequency windows have been chosen as they correspond
to particular cases of interest:
\begin{enumerate}
  \addtolength{\itemsep}{-0.15\baselineskip}
\item[$i$] Photons from the central peak.
\item[$ii$] Case where~$g^{(2)}_\Gamma(0)=2$ (usually attributed to thermal light).
\item[$iii$] Photons from leapfrog emission.
\item[$iv$] Case where~$g^{(2)}_\Gamma(0)=1$ (usually attributed to coherent or uncorrelated light).
\item[$v$] Photons from a side peak.
\end{enumerate}
The central peak is partially thermalized, with a $g^{(2)}(\tau)$ that
closely resembles the form of thermal fluctuations,
$g^{(2)}(\tau) = 1 +\exp(-2|\tau|/\tau_0)$. Upon closer inspection,
however, this is an approximation as the exact solution presents small
departures, in particular, a differentiable slope at the origin and
small ripples that are thinly visible on the theory curve, that we
keep separate from the Monte Carlo data for clarity (the quality of
their agreement is shown in the rightmost column).  Note that the
dynamics of coherent driving of a two-level system is considerably
more complicated than its incoherent counterpart and we could not
find, so far, a general closed-form expression for
$g^{(2)}_\Gamma(\tau)$ in this case for arbitrary frequencies.
Applying the technique of effective-quantum state reconstruction from
the correlators, described in Section~\ref{sec:Sun30Apr113440BST2017},
we find that the statistics~$p(n)$ fits well with a cothermal
distribution with $\approx80\%$ of thermal emission and~$\approx20\%$
of uncorrelated emission. Overall, the emission of the central peak is
thus well described by a mixture of thermal and uncorrelated light. It
is, as such, not very interesting per se. Turning now to the second
frequency window, $ii$, which features~$g^{(2)}_\Gamma(0)=2$,
characteristic of thermal emission, one can now see more clearly the
deviation from the thermal paradigm, with bulging and tails deforming
the correlation function. These are well reproduced by the Monte Carlo
statistics and we let the reader decide if their statistical acuity
lets them, on the small sample of clicks reported here, observe
deviations from the thermal paradigm
(cf.~Fig.~\ref{fig:Sat29Apr131334BST2017}$viii$)

The most interesting window, $iii$, lies halfway between the central
and side peak. This is the frequency at which, according to our
interpretation of the theory,~\cite{gonzaleztudela13a} two photons can
make a leapfrog process from the state~$\ket{+}$ in a given manifold
to the state $\ket{-}$ two manifolds below, jumping over the
intermediate manifold, cf.~the rightmost transition in
Fig.~\ref{fig:Fri26May193749BST2017}. These photons are strongly
correlated in several ways. From a photo-detection point of view, they
should arise as more occurrences of closely-spaced two-photon clicks
than if the emission was uncorrelated. In particular, their rate of
coincidences should increase, leading to $g^{(2)}_\Gamma(0)\gg2$, or
so-called superbunching. This is both predicted by the exact
theory~\cite{delvalle12a,gonzaleztudela13a} and observed in our Monte
Carlo simulations, as seen in
Fig.~\ref{fig:Thu27Apr184800BST2017}. Remarkably, even with as few as
9112 clicks collected in the numerical experiment, we can reconstruct
a high-quality signal, revealing the fine details of its structure.
Note as well that on the real-time series of clicks, out of the nine
photons emitted, four came as two-photon bundles (the fifth and sixth
clicks are so closely spaced as almost overlapping; other ticks are
single-photon events). The small sample of clicks also shows strong
ordering, combining equal spacing and gaps of no emission. While the
latter is characteristic of thermal emission, the former is typically
characteristic of antibunching.  This combination can be seen as the
selection through filtering of strongly correlated emission from the
emitter, rather than tampering from the filters on the statistics:
focusing to this frequency windows allows us to detect the two-photon
emission events that occurs, from the dressed-atom picture, at this
frequency.  It would be rewarding to apply this technique to the
filtered emission of a ``bundler'',~\cite{sanchezmunoz14a} a device
that emits the majority, and in some regime, close to 100\%, of its
light as $N$-photon emission, and for which filtering has been shown
to considerably boost the purity of the quantum
emission.~\cite{thesis_sanchezmunoz16a, arXiv_sanchezmunoz17a} Also
further photon-counting characterization would certainly be
enlightening, and preliminary investigations show that the percentage
of closely-spaced photons is over an order of magnitude higher
in~$iii$ than in the others at the exception of~$ii$, as compared to
which it is only about~$3.8$ times larger. We leave further
characterizations for future works, but provide a last compelling
manifestation of leapfrog emission from the effective quantum state
reconstruction approach,
cf.~Section~\ref{sec:Sun30Apr113440BST2017}. This highlights the
frequency window~$iii$ as the most dissimilar one as compared to the
others, featuring a neat kink at the probability to have two photons,
$p(2)$, showing the relative predominance of two-photon
emission. Overall, this simulation makes it obvious that the emission
in this frequency window suffers from no artifact of post-selection or
normalization, but does indeed provide strongly correlated photon
streams.

The fourth frequency window, $iv$, chosen for its
$g^{(2)}_\Gamma(0)=1$ of uncorrelated emission, is also a case that
shows strong departures at nonzero~$\tau$ due to filtering. This is,
here again, well captured by the Monte Carlo clicks and is visibly
noticeable on the small sample, that features ordered clumps of
uncorrelated clicks. With the last window, $v$, we come back to a case
well studied in the literature, of antibunched emission, albeit far
from perfect ($g^{(2)}_\Gamma(0)\approx 0.42$ and
$\min_\tau g^{(2)}_\Gamma(\tau)\approx 0.37$). The fact that the
minimum antibunching is not at zero is another manifestation of
frequency filtering, thinly visible on the figure as small
oscillations, but not reproduced at this level of signal by the Monte
Carlo data. Correspondingly, the $p(n)$ shows increasingly suppressed
probabilities to get higher number of photons.

\subsection{Cross-correlations}

\begin{figure}[t]
  \centering
  \includegraphics[width=\linewidth]{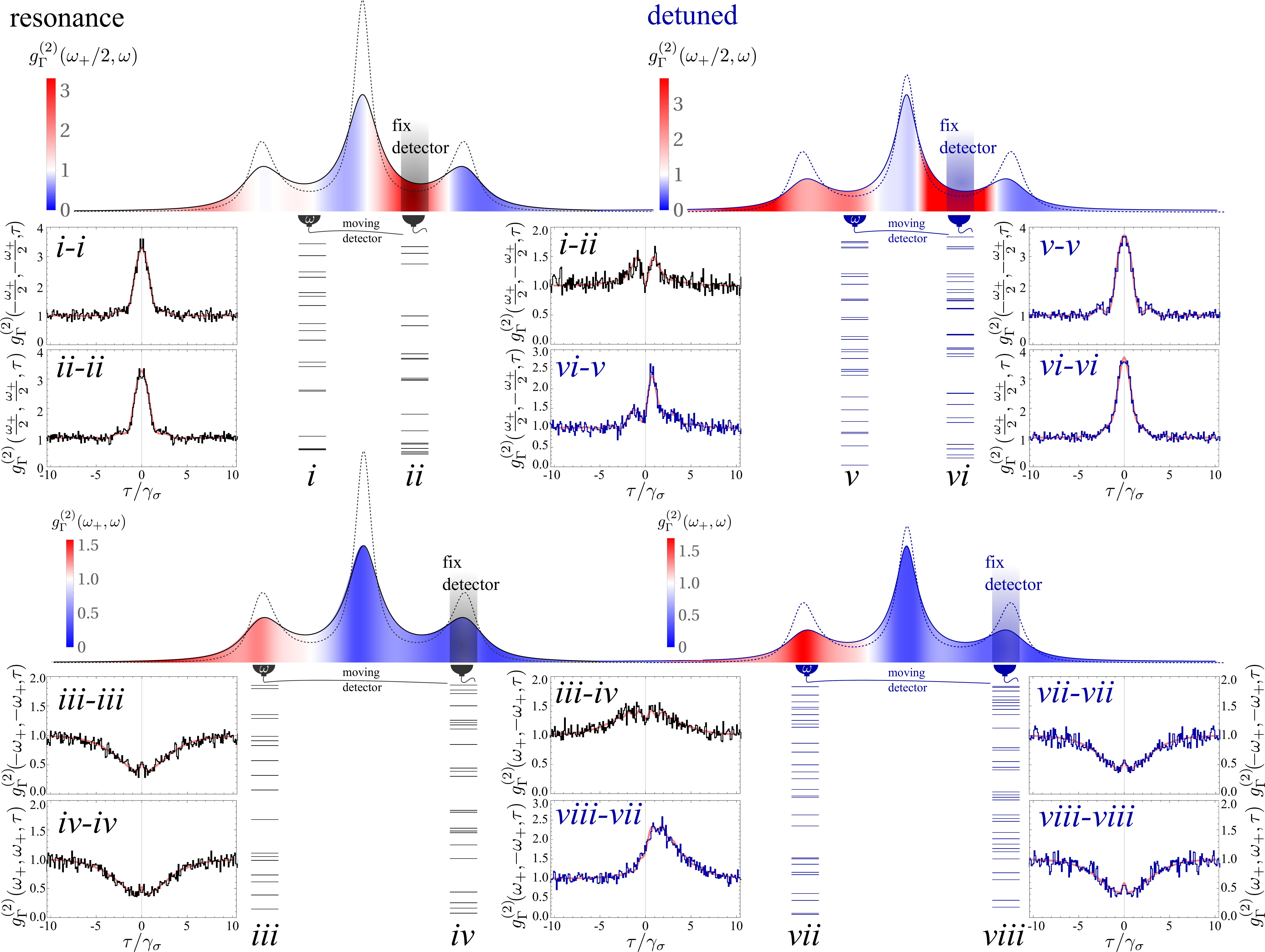}
  \caption{Frequency-resolved Monte Carlo simulation of the Mollow
    triplet in dual frequency windows. The case on the left (right) is
    for driving the two-level system at resonance (with detuning
    $\Delta=1.5\gamma_\sigma$).  The dotted (solid) lineshape is the
    triplet as detected by an ideal (finite-bandwidth~$\Gamma$)
    detector. The color code within the spectra is for one filter kept
    fixed at a leapfrog window (top case) or at a sideband (bottom
    case). Sequences of clicks have been recorded in the windows $i$
    and~$ii$ and $iii$ and~$iv$, the two groups being independent
    (that is, clicks between, e.g., $i$ and~$iii$ are not
    correlated). Small samples are displayed, with renormalization of
    time to compare equal intensities between the two groups (time is
    the same within each group).  Strong correlations of photons with
    different frequencies are clear, in particular, the simultaneity
    of leapfrog emission and their heralding character with detuning
    are obvious. The cross-correlations are shown as measured by the
    Monte Carlo data (left column) and as computed by the theory of
    frequency-resolved photon correlation (2nd column,
    red). $1/\gamma_\sigma$ sets the unit and $\Omega=5\gamma_\sigma$,
    $\Delta=1.5\gamma_\sigma$.}
  \label{fig:Thu29Jun112024BST2017}
\end{figure}

In this final part, we consider cross-correlations, for which the
Mollow triplet is also a particularly suitable lineshape. That is to
say, we consider two detectors acquiring data simultaneously. The
master equation for two detectors upgrade
Eq.~(\ref{eq:SunApr23165942BST2017}) to:
\begin{equation}
  \label{eq:Mon6Nov222642GMT2017}
  \partial_t \rho = i [\rho,H_\sigma+\sum_{i=1,2}\omega_{\xi_i} \ud{\xi_i}\xi_i] +
  \frac{\gamma_\sigma}{2} \mathcal{L}_\sigma \rho +
  \sum_{i=1,2}\left( \frac{\gamma_{\xi_i}}{2}  
    \mathcal{L}_{\xi_i} \rho + \sqrt{\alpha_i \gamma_\sigma \gamma_{\xi_i}} \lbrace
    [\sigma \rho, \ud{\xi_i}] + [\xi_i,\rho\ud{\sigma}] \rbrace \right)
\end{equation}
with~$\xi_1$, $\xi_2$ the two detectors.The factors
$\alpha_1=(1-\chi_0-\chi_1)(1-\chi_2)$ and
$\alpha_2=\chi_0(1-\chi_3)$, satisfying simultaneously
$0\leq \chi_0,\chi_1,\chi_2,\chi_3\leq1$ and $\chi_0+\chi_1\leq1$,
take into account the several decay channels of the source: a
fraction~$\chi_1$ into free space, a fraction~$\chi_0$ to the
detector~$\xi_1$ and the remaining fraction~$(1-\chi_0-\chi_1)$ to the
detector~$\xi_2$. In analogy with the case of a single detector, the
system described by the master
equation~(\ref{eq:Mon6Nov222642GMT2017}) has five collapse operators:
\begin{subequations}
  \begin{align*}
  \label{eq:FriNov10190141CET2017}
    \quad c_1 = \sqrt{\chi_0 \gamma_\sigma}\,\sigma +
 \sqrt{(1-\chi_2) \gamma_{\xi_1}}\,\xi_1 \,, \quad
  c_2 = &\sqrt{(1-\chi_0-\chi_1) \gamma_\sigma}\,\sigma +
 \sqrt{(1-\chi_3) \gamma_{\xi_2}}\,\xi_2 \,, \quad
 c_3 = \sqrt{\chi_1 \gamma_\sigma} \,\sigma\,, \quad\\
 c_4 = \sqrt{\chi_2 \gamma_{\xi_1}}& \,\xi_1\, \quad
 \mathrm{and} \quad
    c_5 =\sqrt{\chi_3 \gamma_{\xi_2}}\, \xi_2\,,
    \end{align*}
\end{subequations}
and its associated non-hermitian Hamiltonian becomes
\begin{equation}
  \label{eq:FriNov10192531CET2017}
  \tilde H= H_\sigma+H_{\xi_1}+H_{\xi_2} -  i (\sqrt{\alpha_1 \gamma_\sigma
               \gamma_{\xi_1}}\, \ud{\xi_1}\sigma + \sqrt{\alpha_2
               \gamma_\sigma \gamma_{\xi_2}}\, \ud{\xi_2}\sigma)-
             \frac{i}{2}( \gamma_\sigma \ud{\sigma}\sigma + \gamma_{\xi_1}
             \ud{\xi_1}\xi_1 + \gamma_{\xi_2}\ud{\xi_2}\xi_2 ) \,.
\end{equation}
As for the case of autocorrelations, one could similarly demonstrate
the equivalence between cross-correlations to any orders as computed
through the frequency-resolved photon correlations and those obtained
through Eq.~(\ref{eq:Mon6Nov222642GMT2017}) above. Also as was done
before for single frequency windows, by applying the Monte Carlo
techniques to the detectors, one can thus obtain simulated photon
emissions, this time in two frequency windows. Computing the
correlations from this raw data provides a numerical version of the
theoretical correlations.  This is shown in
Fig.~\ref{fig:Thu29Jun112024BST2017} for the joint emission of the two
sidebands on the one hand, and then of the two leapfrog windows on the
other hand, both when driving the two-level system at resonance or
with a detuning.

While we considered a small subset only of the possible
autocorrelations in Fig.~\ref{fig:Thu27Apr184800BST2017} for the Monte
Carlo data, we could still provide a comprehensive theoretical result
at least for~$g^{(2}_\Gamma(0)$ through the color-coded spectrum. For
cross-correlations, however, this would require a 2D plot to reproduce
the entire two-photon correlation spectrum.~\cite{gonzaleztudela13a}
Instead, we consider here the case where one detector is fixed and the
other one sweeps the rest of the spectrum, providing the
cross-correlations. We then place the other detector for Monte Carlo
sampling at the location of interest. As before, we show raw photon
emission, but with no time-rescaling as the respective frequencies
chosen have similar intensities. We also compare two-photon
correlations computed from this data (black and blue lines, at
resonance and with detuning respectively) with the theoretical result
(red lines). The main difference between these cases and the previous
ones is that the two streams of photons are now correlated as the
detectors are measuring simultaneously. If restricting to one stream
only, we recover the previous cases, so in
Fig.~\ref{fig:Thu29Jun112024BST2017}, panels~$i$-$i$ and $ii$-$ii$ on
the one hand, as well as $iii$-$iii$ and $iv$-$iv$ on the other hand,
can be found in Fig.~\ref{fig:Thu27Apr184800BST2017}, in panels~$iii$
and~$v$ respectively. The two cases have different parameters, since
with two detectors, the simulation is more computer intensive and we
chose a case that provides more emission between the peaks. One can
check however how the qualitative shapes of the correlations remain
the same.  At resonance, both streams provide the same type of
correlations, but with detuning, they could be different. This is
indeed the case for the leapfrog emission, and small departures can be
observed between $v$-$v$ and $vi$-$vi$, both in the theoretical line
and the Monte-Carlo generated data: the oscillations are more marked
for the low-energy window and a depletion is indeed visible
in~$vi$-$vi$ that disappeared in~$v$-$v$. The side peaks, however,
feature similar correlations. This shows again the typical richer
dynamics away from the peaks.

Both at resonance or with detuning, what is of interest when detecting
in different windows simultaneously is their cross-correlations, as
shown in the central column of Fig.~\ref{fig:Thu29Jun112024BST2017}
with panels $i$-$ii$, $vi$-$v$, $iii$-$iv$ and $viii$-$vii$. There are
now clear features in these cross-correlations, whereas the same
procedure applied to the streams of the previous cases, features no
correlations, i.e., one obtains flat lines. At resonance, the
cross-correlations are symmetrical in time. The side peaks
correlations feature tiny oscillations which are however too small to
be observed with the amount of signal we acquired in the numerical
experiment, and they are hidden by its fluctuations. With detuning,
they can and do become time-asymmetrical, as shown in panels~$vi$-$v$
and~$viii$-$vii$. In such cases, the order of detection matters, and
in both cases, the detection on first detector, $vi$ or $viii$,
respectively, makes it more likely to later detect (with
around~$1/\gamma_\sigma$ delay) a photon on the second detector, $v$
and $vii$, respectively. The strength of such correlations, less
than~3, is still fairly modest to call this heralding, but this is the
basis for such a mechanism to be exploited with proper engineering,
such as Purcell-enhancement.  Like before, our procedure yields
correlated streams of photons of different frequencies, that we have
just shown through their agreement with the theory of
frequency-resolved photon correlations, simulate the actual photon
emission from the system. One could use this raw data to compute
numerically, e.g., counting or time-delay distributions, otherwise not
easily accessible theoretically.  Of course, the scheme could in
principle be extended to any number of detectors and allow
consequently higher orders of correlation to be computed in this
way. In the limit of an infinite number of detectors, each with a
given frequency and vanishing spectral width, one would thus simulate
the ideal emission of the system. With a finite number of detectors
with a finite bandwidth, one would simulate its filtered emission. We
believe that a complexity analysis of the correlations would allow to
use the emission of the two-level system as a simpler platform than
boson sampling~\cite{aaronson11a} to test quantum supremacy by making
a laboratory measurement which no classical computer would be able to
simulate.

\section{Conclusions and perspectives}

In conclusion, we have presented computer experiments that simulate
numerically the photon emission from a quantum emitter, specifically,
a two-level system under both coherent and incoherent driving, at both
low and large pumping. Our approach is based on the Quantum Monte
Carlo technique~\cite{zoller87a,carmichael89b,dalibard92a,molmer96a}
applied to the cascaded formalism.~\cite{gardiner93a,carmichael93a} We
have shown how the correlations computed from the raw data of the
simulation match with the theoretical results provided by the theory
of frequency-resolved photon correlations.~\cite{delvalle12a} In the
simplest case, we have shown how filtering spoils antibunching and
turns a two-level system into a thermal source, albeit in a more
subtle way than is usually assumed and for which we have provided
exact closed-form expressions.  We have also shown more generally how
frequency filtering provides a resource to tailor and engineer photon
statistics, in particular thanks to its selection of strongly
correlated processes such as ``leapfrog'' transitions that consist in
the simultaneous emission of a photon bundle between two
non-contiguous dressed states in the level structure of the
system.~\cite{gonzaleztudela13a}  This displays rich and potentially
useful features that are captured in the Monte Carlo simulation and
that would be similarly observed experimentally.  An apparent
shortcoming is that the signal is scarce in frequency windows that are
the most strongly correlated. This is however a direct consequence of
dealing with the quantum part of the signal: there is less of
it. Frequency filtering acts as a process akin to distillation, with
the same consequence of providing quality at the expense of
quantity.~\cite{delvalle13a} Nevertheless, quantum engineering can
come at the rescue and already the oldest trick of cavity
QED---Purcell enhancement---allows, in some regime, to have \emph{all}
the light of the system emitted as strongly correlated
photons.~\cite{sanchezmunoz14a} Using cavities to Purcell-enhance
leapfrog processes, one can devise new generations of heralded
$N$-photon sources, or, even more generally, bring the system to emit
in any desired distribution of
photons.~\cite{lopezcarreno17a} Such configurations remain to be
studied in detail and, of course, implemented in the laboratory.  This
should provide one route for universal multi-photon sources, with
heralded~$N$ photon sources as the most elementary realization. Since
leapfrog processes are energy-conserving $N$-photon relaxations, they
also appear particularly suitable for energy-time entanglement
emission, that power a class of quantum-cryptographic protocols with
technical advantages as compared to those based on entangling in
polarization.  The latest work from Peiris \emph{et
  al.},~\cite{peiris17a} who is so far leading the laboratory
implementation of this emerging branch of quantum optics, focused on
the side peaks emission and, as a consequence, failed to break the
barrier of a Bell violation. This has been argued no to be a proof of
nonlocality anyway~\cite{arXiv_jogenfors17a} due to its 50\%
post-selection.~\cite{aerts99a} It is easily computed that leapfrog
emission would break the Franson limit, but in the light of the
Franson configuration's loophole,~\cite{jogenfors15a} the new
challenge is to turn to stricter conditions of Bell violations such as
Chained Bell's inequalities. While this has been recently
demonstrated,~\cite{tomasin17a} the tunable statistics from the Mollow
triplet and its windows of strong correlations make it a promising
platform to further test and advance this line of research.  Finally,
the combinatoric aspects that quickly make such simple problems
numerically forbidding also suggest that a two-level system could be
used in the laboratory for tests of quantum supremacy directly from
photon detections, without a complex system of beam splitters
intervening to bring in the quantum
complexity.~\cite{broome13a,crespi13a} All these results leave much
for room for further works, and we foresee that frequency-resolved
photon correlations will become a major theme of photonics. They are
relevant even when they are ignored and awareness of the underlying
physics should allow to considerably optimize, tune and expand the
range of applications of quantum light sources.

\section*{Acknowledgments}

Funding by the Newton fellowship of the Royal Society, the POLAFLOW
ERC project No.~308136, the Ministry of Science and Education of
Russian Federation (RFMEFI61617X0085), the Spanish MINECO under
contract FIS2015-64951-R (CLAQUE) and by the Universidad Aut\'onoma de
Madrid under contract FPI-UAM 2016 is gratefully acknowledged.

\section*{Author contributions statement}

EdV an JCLC developed the theory, JCLC implemented and performed the
numerical simulations, FPL and JCLC composed the figures, FPL wrote
the text and supervised the research.  All the authors conceived the
work, discussed the results and contributed to the manuscript.

\section*{Competing Financial Interests statement}

The authors declare no competing financial interests.

%\section*{References}

% References
\bibliography{Sci,arXiv,books} % bibliography data in report.bib

\end{document}